\documentclass[mnsc,nonblindrev]{informs3} 

\OneAndAHalfSpacedXI

\usepackage{natbib}
 \bibpunct[, ]{(}{)}{,}{a}{}{,}%
 \def\bibfont{\small}%

\usepackage{amsmath,amssymb,amsfonts,bbm}
\usepackage{enumerate,ifthen,comment,xcolor,graphicx}
\graphicspath{ {./}{../} }

\TheoremsNumberedThrough     
\ECRepeatTheorems

\EquationsNumberedThrough    

\MANUSCRIPTNO{MS-17-01173}

\begin{document}


\RUNAUTHOR{Gopalakrishnan, Mukherjee, and Tulabandhula}

\RUNTITLE{The Costs and Benefits of Sharing}

\TITLE{The Costs and Benefits of Sharing: Sequential Individual Rationality and Fairness}

\ARTICLEAUTHORS{%
\AUTHOR{Ragavendran Gopalakrishnan}
\AFF{Conduent Labs India (formerly Xerox Research Centre India), \EMAIL{Ragavendran.Gopalakrishnan@conduent.com} \URL{}}
\AUTHOR{Koyel Mukherjee}
\AFF{IBM Research India, \EMAIL{koyelmjee@gmail.com} \URL{}}
\AUTHOR{Theja Tulabandhula}
\AFF{University of Illinois Chicago, \EMAIL{tt@theja.org} \URL{}}
} 

\ABSTRACT{%
In designing dynamic shared service systems that incentivize customers to opt for shared rather than exclusive service, the traditional notion of individual rationality may be insufficient, as a customer's estimated utility could fluctuate arbitrarily \textit{during} their time in the shared system, as long as their realized utility at service completion is not worse than that for exclusive service. In this work, within a model that explicitly considers the ``inconvenience costs'' incurred by customers due to sharing, we introduce the notion of \textit{sequential} individual rationality (SIR) that requires that the \textit{disutility} of existing customers is nonincreasing as the system state changes due to new customer arrivals. Next, under SIR, we observe that cost sharing can also be viewed as \textit{benefit sharing}, which inspires a natural definition of \textit{sequential} fairness (SF)---the total incremental benefit due to a new customer is shared among existing customers in proportion to the incremental inconvenience suffered.

We demonstrate the effectiveness of these notions by applying them to a ridesharing system, where unexpected detours to pick up subsequent passengers inconvenience the existing passengers. Imposing SIR and SF reveals interesting and surprising results, including: (a)~natural limits on the incremental detours permissible, (b)~exact characterization of ``SIR-feasible'' routes, which boast sublinear upper and lower bounds on the fractional detours, (c)~exact characterization of sequentially fair cost sharing schemes, which includes a strong requirement that \textit{passengers must compensate each other for the detour inconveniences that they cause}, and (d)~new algorithmic problems related to and motivated by SIR.
}%


\maketitle

\section{Introduction}

Sharing of resources and services is ubiquitous in today's economy, driven by increasing costs to the individual of enjoying exclusive access. For example, increased congestion on roads has pushed more commuters towards public transportation and ridesharing~\citep{mckenzie2015drives,sivak2016recent}. Funding shortages of governments and thin profit margins of businesses force more people to share a smaller set of customer-serving resources both physically (contact centers~\citep{aksin2007modern}, airport security checks~\citep{cole2015improving,zou2015us}) and virtually (online services~\citep{armbrust2010view}). The popularity of cloud computing services (e.g., Amazon Web Services, Microsoft Azure) has increased because they drastically bring down the computing infrastructure costs~\citep{kondo2009cost,ahn2012dynamic}. Even more examples include spectrum sharing in wireless networks~\citep{peha2009sharing}, and shared logistics in supply chain distribution networks~\citep{bowersox2000ten}.

Of particular interest are shared service systems where arriving customers spend a finite amount of time and money in the system getting served, and leave the system upon service completion. However, because the service is shared, the time taken for a customer to be served, as well as the monetary cost of service, can change depending upon the arrival/departure of other customers into/from the system. Usually, the more the customers that share the same amount of resources, the more time and less money each individual customer spends in the system. In ridesharing, picking up an additional passenger involves a detour which increases passengers' commute times, but brings down their shares of the total cost. In a priority queue, there are multiple service levels; those with higher priority (and hence, shorter average waiting times) cost more~\citep{Katta2005}. Due to resource pooling in cloud computing services, the response times for the same load can be higher if other tenants sharing the same set of resources place concurrent loads, and so, offerings with such performance variability are priced lower~\citep{Jackson2011}.

When designing the pricing or cost sharing scheme for a shared service system, two major factors that influence a customer's choice of shared service must be considered: (a)~Individual Rationality (IR), where a customer compares her own utility between the shared and exclusive service options, and asks herself if the cheaper monetary cost of the shared service is worth the longer waiting/response time, and (b) Fairness, where a customer compares her utility with those of other customers in the system, and evaluates whether everyone is ``equitably'' better off in opting for the shared service. While traditional notions of IR and fairness capture the requirement that customers who opt for a shared service are ``happier'' than they would be had they availed an exclusive service, they fail to address a customer's experience \textit{during} the time they are in the system due to subsequent arrivals (which could be unexpected), e.g., a sudden burst of arrivals into a higher-priority queue could frustrate a customer waiting in a lower-priority queue at that moment, even if she is later offered a discount for having waited longer. In ridesharing, a frequent source of frustration are the (often unexpected) detours taken to pick up and/or drop off additional passengers, which inconvenience existing passengers.

\textit{Our goal in this paper is to address this concern by defining appropriate new notions of IR and fairness, and demonstrate their effectiveness by investigating their consequences (characteristic and algorithmic) within the context of a specific application, namely ridesharing.}

\subsection{Our Contributions}

First, in Section~\ref{sec:sir-sf-gen}, we introduce the concepts of \textit{sequential} IR (SIR) and \textit{sequential} fairness (SF), that extend their traditional counterparts to be applicable at a finer granularity to a dynamic shared service system. This involves invoking appropriate IR and fairness constraints every time the state of the system changes due to a new customer arrival.\footnote{The concepts can be naturally extended to other situations which affect the customer experience, e.g., adding/removing service capacity, customer departures, etc.} We model the disutility of a customer as the sum of the monetary cost of service and an ``inconvenience cost'' due to the presence of other customers in the system. While IR simply requires the disutility when opting for the shared service to be not greater than that for an exclusive service, SIR additionally requires the disutility to be nonincreasing throughout the time spent in the shared system. The companion fairness concept, SF, requires that the marginal decrease in disutility (the \textit{benefit} of sharing) every time the state of the system changes is ``equitably'' experienced by all the customers in the system. We also briefly discuss a concern that these strong concepts could be too restrictive to allow feasible practical policies in certain scenarios.

Next, we apply these concepts to a ridesharing system (motivated by its impact on promoting sustainable behavior), where cost sharing among passengers is a major design component.\footnote{Other elements are effective ride matching, routing, and, in the case of commercial ridesharing, profitability; we explain how our focus on the costs influences these other factors, e.g., by adding the SIR and SF constraints to a matching/routing algorithm that is already in use.} SIR targets the oft-lamented pain points due to detours experienced by passengers during the ride, by ensuring that existing passengers are progressively better off every time an additional passenger is picked up. In addition, SIR also ensures a certain degree of robustness, e.g., in a dynamic/online setting, passengers would remain satisfied even if a future pickup is canceled. We show that imposing SIR and SF on the routing and cost sharing schemes of a ridesharing system is not restrictive; to the contrary, it brings out several interesting and surprising consequences:
\begin{enumerate}[(a)]
\item In Section~\ref{sec:sir}, we provide an exact characterization (Theorem~\ref{THM:SIR-FEASIBILITY}) for any route to be ``SIR-feasible'', that is, there exists some budget-balanced cost sharing scheme that is SIR on that route. These SIR-feasibility constraints, though fairly straightforward to derive, are necessarily complex, so in Section~\ref{SEC:SINGLE}, we consider a simplified scenario where all the passengers are travelling to a common destination. For this ``single dropoff'' scenario, we show that the SIR-feasibility constraints simplify to \textit{natural} upper bounds on the incremental detours, that keep shrinking as the ride progresses towards the destination and as more passengers are picked up.
\item In a series of theorems (Theorems~\ref{thm:SIR-starvation-no-detour}-\ref{thm:SIR-starvation-lower}) in Section~\ref{sec:starvation-bounds}, we show that the above bounds on incremental detours can be aggregated to establish upper and lower bounds on the total detour endured by a passenger as a fraction of their shortest distance to their destination. These bounds depend on how sensitive the passengers are to detours; in realistic scenarios, these bounds are \textit{sublinear} in the number of passengers.
\item In Section~\ref{sec:sequential-fairness}, we present an exact characterization of sequentially fair cost sharing schemes for the single dropoff scenario (Theorem~\ref{thm:sf-char}), which exposes several practical structural properties of such schemes, including a surprisingly strong requirement that \textit{passengers must compensate each other for the detour inconveniences that they cause}.
\item Finally, in Section~\ref{SEC:ALG-SIR}, we explore some important algorithmic questions motivated by SIR. In particular, it is unknown, even for the single dropoff scenario, whether there exists a polynomial time algorithm to check for the existence of SIR-feasible routes, when restricted to an arbitrary metric space (we show that it is NP-hard otherwise). Even if so, we show that optimizing for total distance traveled over SIR-feasible routes is NP-hard (through a reduction from a variant of Metric-TSP). We then consider a variant of the vehicle routing problem where passengers are allocated to vehicles such that the total vehicle-miles traveled is minimized. While this problem is known to be NP-hard in general, we show that it can be solved in polynomial time given a fixed ordering on the pickup points.
\end{enumerate}

\subsection{Related Work}

The cost sharing problem for ridesharing has garnered relatively little attention in literature (compared to the ride matching and route optimization problems)---in most existing schemes, individual passengers are asked to post what they are willing to pay in advance~\citep{sharek}, share the total cost proportionately according to the distances travelled~\citep{Geisberger2010,Agatz2011}, or negotiate their cost shares on their own during/after the ride. Such methods ignore the real-time costs and delays incurred during the ride (as in the first instance), are insensitive to the disproportionate delays encountered during the ride (as in the second instance), or lead to a complicated and often uncomfortable negotiation process between possible strangers (as in the third instance).

Recent work has studied cost sharing when passengers have significant autonomy in choosing rides or forming ridesharing groups, e.g., cost sharing schemes based on the concept of kernel in cooperative game theory~\citep{bistaffa}, second-price auction based solutions~\citep{Kleiner2011}, and market based ride-matching models with deficit control~\citep{zhao}. Fair cost sharing in ridesharing has also been studied under a mechanism design framework by~\cite{kamar}, where an individually rational VCG-based payment scheme is modified to recover budget-balance at the cost of incentive compatibility, and by~\cite{nguyen}, where customers are offered an additive, detour-based discount, and the allocations and pricing are determined through an auction. Our work differs from all the above in that we do not make any assumptions about the mechanics of ride matching; our cost sharing model is independent of the routing framework (static or dynamic), and is applicable to community carpooling and commercial ridesharing alike. Moreover, we define a monotonic form of individual rationality for dynamic ridesharing.

Our work is different from the problem of pricing in ridesharing (see, e.g.,~\cite{Banerjee2015,Bimpikis17}); our focus is on sharing the resulting cost among the passengers.

Previous works on ridesharing that address individual rationality and detour limits treat them as \textit{independent} constraints, e.g.,~\cite{kamar,Santos2013,Pelzer2015}. In contrast, in our model, requiring (sequential) individual rationality \textit{induces} natural bounds on detours experienced by the ridesharing passengers.

There is a plethora of work when it comes to optimization problems in ridesharing~\citep{Agatz2012,Furuhata2013,Pelzer2015,ozkan2016dynamic,alonso2017demand}. While the detour constraints that SIR induces can augment any routing optimization problem, in this paper, we focus on finding an allocation of passengers to vehicles that minimizes the total vehicle-miles, which is a variant of the vehicle routing problem~\citep{Cordeau2007}, with the additional constraint of SIR.

Variations of individual rationality involving temporal aspects are well studied in the economics literature, e.g., \textit{ex-ante}, \textit{interim}, and \textit{ex-post} individual rationality in mechanism design~\citep{Narahari2009}, and sequential individual rationality in bargaining and repeated games~\citep{Esteban1991}. However, we are the first to explore its applicability to dynamic shared service systems and fairness properties of the resulting outcomes.

An extensive literature on cooperative game theory and fair division~\citep{Moulin2004,Jain2007} offers various cost sharing schemes that can be analyzed in our framework. Our view of fairness relies on how the total incremental benefit due to ridesharing is allocated among the passengers during each stage of the ride (sequential fairness). While we believe the two approaches are not independent, exploring the connections is beyond the scope of this work.

We now outline some of the related work in cost sharing in a dynamic shared resource allocation setting, in which our contributions can potentially discover alternate solutions with different desiderata. For instance,~\cite{elmachtoub2014cost} study online allocation problems where customers arrive sequentially, and decisions regarding whether to accept or reject customers must be made upon their arrival, with the goal of minimizing the sum of production costs (for accepted customers) and rejection costs.

Additionally, the trade-off of (general forms of) fairness with efficiency for allocation problems (in different domains, ranging from online advertisement portals~\citep{bateni2016fair} to manufacturing and retail~\citep{haitao2007fairness}) has been widely studied in the literature, and recently conceptualized and studied as the ``price of fairness''~\citep{bertsimas2011price,dickerson2014price,heydrich2015dividing}. We hope that the concepts of individual rationality and fairness that our work introduces would inspire similar studies (theoretical and empirical) on their trade-offs with different notions of efficiency such as social welfare maximization and profit optimization.

\section{Sequential Individual Rationality and Sequential Fairness}\label{sec:sir-sf-gen}

In this section, we formally define the notions of sequential individual rationality and sequential fairness for shared service systems. We introduce the necessary notation first.

Let $\mathcal{N}=\{1,2,\ldots,n\}$ denote the set of customers, ordered according to their arrival times. Let $t_i>0$ denote the time at which customer $i\in\mathcal{N}$ arrives into the system, and let $T=\{t_0,t_1,t_2,\ldots\}$, where $t_0=0$. Let $\ell(i)$ denote the last customer to arrive into the system before $i$ leaves the system. Let $s_u$ denote the system state at time $u$, which encodes the necessary information about all the customers that are in the system at time $u$. Let $S_{ij}=\{s_u\ |\ u\in T\ \mbox{and}\ t_i\leq u\leq t_j\}$ denote the information set of states from time $t_i$ to $t_j$.

For any $j\in\mathcal{N}$, let $S(j)=\{1,2,\ldots,j\}$ denote the set of customers who have arrived until $t_j$, and let $\mathcal{OC}(S_{1j})$ denote the operating cost of the system, conditional on there being no more arrivals after $t_j$. $f$ denotes the cost sharing scheme according to which the operating cost is shared among the customers. In particular, the monetary cost of service to customer $i\in S(j)$, conditional on there being no more arrivals after $t_j$, is $f(i,S_{ij})$.

\begin{definition}
A cost sharing scheme $f$ is \textit{budget balanced} if
\begin{equation}\label{eq:BB-gen}
\sum_{i\in S(j)} f(i,S_{ij}) = \mathcal{OC}(S_{1j})\qquad\forall\ j\in\mathcal{N}.
\end{equation}
\end{definition}

For any $j\in\mathcal{N}$, the inconvenience cost incurred by customer $i\in S(j)$ due to all the other customers she encounters in the system, assuming no customers arrive after time $t_j$, is denoted by $\mathcal{IC}_i(S_{ij})$. Several factors may affect the monetary and inconvenience costs, whose exact functional forms would depend on the mechanics of the system being modeled.\footnote{For example, one way to model the inconvenience cost could be to measure the additional time the customer spends in the system compared to when she is the sole customer served, scaled by how much she values a unit of her time.}

\subsection{Disutility and Individual Rationality (IR)}

The \textit{disutility} of a customer $i$, assuming no customers arrive after time $t_j$, is defined as the the sum of their monetary and inconvenience costs, that is,
\begin{equation}\label{eq:DU-gen}
\mathcal{DU}_i(t_j) =
\begin{cases}
\mathcal{DU}_i^0, & 0\leq j < i\\
f(i,S_{ij}) + \mathcal{IC}_i(S_{ij}), & i\leq j\leq \ell(i)\\
\mathcal{DU}_i(t_{\ell(i)}), & j > \ell(i)
\end{cases}
\end{equation}
where $\mathcal{DU}_i^0$ denotes the disutility corresponding to the exclusive service to customer $i$.

\begin{definition}
A cost sharing scheme $f$ is \textit{Individually Rational} (IR) if
\begin{equation}\label{eq:IR-gen}
\mathcal{DU}_i(t_n) \leq \mathcal{DU}_i(t_0)\qquad\forall\ i\in\mathcal{N}.
\end{equation}
\end{definition}

\subsection{Sequential Individual Rationality (SIR)}

While IR ensures that a customer's disutility at the time of service completion in a shared system is not greater than that of an exclusive service, it still allows for the disutility to fluctuate arbitrarily \textit{during} the time spent in the system, which can negatively affect the customer experience.

\begin{definition}
A cost sharing scheme $f$ is \textit{Sequentially Individually Rational} (SIR) if
\begin{equation}\label{eq:SIR-gen}
\mathcal{DU}_i(t_j) \leq \mathcal{DU}_i(t_{j-1}) \qquad\forall\ 1\leq j \leq n\qquad\forall\ i\in\mathcal{N}.
\end{equation}
\end{definition}

\subsection{The Benefit of Sharing and Sequential Fairness}\label{sec:sequential-fairness-gen}

Under a cost sharing scheme that is IR, the decrease in disutility to a customer due to her participation in a shared service system (the difference between the right and left hand sides of her IR constraint~\eqref{eq:IR-gen}) can be viewed as her benefit of sharing. Further, it can be seen that the total benefit of sharing, obtained by summing the individual benefits, is independent of the cost sharing scheme, as long as it is budget-balanced. This observation exposes an underlying ``duality'' -- a cost sharing scheme can, in fact, be viewed as a \textit{benefit sharing scheme}. Such a view invites defining cost sharing schemes based on traditional notions of fairness, e.g., a fair cost sharing scheme should distribute the total benefit among the service-sharing customers suitably proportionately.

We extend this notion to budget balanced cost sharing schemes that are SIR by looking into how they distribute the total \textit{incremental} benefit due to each subsequent customer arriving into the system, leading to a natural definition of \textit{sequential fairness}.

\begin{definition}
The \textit{incremental benefit} to customer $i\in\mathcal{N}$ due to the arrival of an incoming customer $j\in\mathcal{N}$ is given by
\begin{equation}\label{eq:IB-gen}
\mathcal{IB}_{i}(S_{ij}) = \mathcal{DU}_i(t_{j-1}) - \mathcal{DU}_i(t_j).
\end{equation}
\end{definition}

\begin{definition}
The \textit{total incremental benefit} due to the arrival of an incoming customer $j\in\mathcal{N}$ is given by
\begin{equation}\label{eq:TIB-gen}
\mathcal{TIB}(S_{ij}) = \sum_{i\in S(j)}\mathcal{IB}_{i}(S_{ij}).
\end{equation}
\end{definition}

We take a very general, but minimal, approach to defining sequential fairness. All that is required of a cost sharing scheme to be sequentially fair is that, when an incoming customer $j$ arrives into the system, the portion of the total incremental benefit that is enjoyed by a previous customer $i$ ($1\leq i\leq j-1$), is proportional to the incremental inconvenience cost to $i$ due to the incoming customer. This is formalized in the following definition.

\begin{definition}\label{def:SF-gen}
Given a vector $\vec{\beta}=\left(\beta_2,\beta_3,\ldots,\beta_n\right)$, where $0\leq\beta_j\leq 1$ for $2\leq j\leq n$, a budget balanced, SIR cost sharing scheme $f$ is \textit{$\vec{\beta}$-sequentially fair} if, for all $2\leq j\leq n$,
\begin{equation}\label{eq:SF-gen}
\frac{\mathcal{IB}_{i}(S_{ij})}{\mathcal{TIB}(S_{ij})} =
\begin{cases}
\beta_j\frac{\mathcal{IC}_i(S_{ij})-\mathcal{IC}_i(S_{i(j-1)})}{\sum_{m=1}^{j-1}\left(\mathcal{IC}_m(S_{mj})-\mathcal{IC}_m(S_{m(j-1)})\right)}, & 1\leq i\leq j-1\\
1-\beta_j, & i=j.
\end{cases}
\end{equation}
\end{definition}

Here, $1-\beta_j$ is the fraction of the total incremental benefit enjoyed by the incoming customer $j$ as a result of joining the service system, and $\beta_j$, the remaining fraction, is split among the previous customers. Setting $1-\beta_j = \frac{\mathcal{IC}_j(S_{jj})}{\sum_{m=1}^{j}\left(\mathcal{IC}_m(S_{mj})-\mathcal{IC}_m(S_{m(j-1)})\right)}$ corresponds to a special case where the incoming customer is treated just the same as everyone else.

\subsection{Weaker Notions of SIR}\label{ssec:weaker-SIR-SF-gen}

There may be situations where SIR is so strong that no feasible practical policy can be expected to satisfy it. For example, in shared service systems where the cost of service is fixed, e.g., priority queues at airports, banks, and hospitals, SIR-compliant policies would require dynamically adding service capacity to counter the inconvenience to low-priority customers every time a high-priority customer arrives into the system. However, it may not be practical to do so each and every time such an arrival occurs. Therefore, weaker notions of SIR could be proposed, e.g., approximate SIR, where only a bounded increase in disutility is allowed (which would induce threshold staffing policies). In addition, when arrivals and service times are modeled probabilistically, as in a queueing system, appropriate probabilistic notions of SIR may be needed, since requiring SIR on every sample path could be too restrictive. Future work should explore such interesting extensions.

\section{A Model for Cost Sharing in Ridesharing}\label{sec:model-ridesharing}

In this section, we instantiate the above general model for a ridesharing system, where $\mathcal{N}$ denotes the set of passengers. For each passenger $i\in \mathcal{N}$, let $S_i$ and $D_i$ denote their pickup and dropoff points, which are assumed to belong to an underlying metric space.

\vspace{0.05in}
\noindent\textbf{Additional Notation:} We assume access to a routing algorithm that, given any subset $S\subseteq \mathcal{N}$, computes a valid route $r_S$ (an ordered sequence of pickup/dropoff points) that serves all the passengers in $S$. Thus, we define the following distance functions for any subset $S\subseteq \mathcal{N}$: (a)~$d(S;r_S)$ denotes the total distance traveled along route $r_S$, and (b)~$d_i(S;r_S)$, for $i\in S$, denotes the total distance traveled along route $r_S$ from $S_i$ to $D_i$. For simplicity, we assume that the costs are completely determined by the traversed distances.\footnote{It is straightforward to extend our model and results to costs that depend on a combination of distance and time.}

Accordingly, the operational cost (or the ``meter fare''), and the inconvenience cost are
\begin{align}
\mathcal{OC}(S;r_S) &= \alpha_{op} d(S;r_S),\ \mbox{and}\label{eq:oc}\\
\mathcal{IC}_i(S;r_S) &= \alpha_i \left( d_i(S;r_S) - d_i(\{i\};r_{\{i\}}) \right),\label{eq:ic}
\end{align}
where $\alpha_{op}>0$ is the price (in commercial ridesharing) or operating cost (in community carpooling) per unit distance, and, for each $i\in \mathcal{N}$, $\alpha_i\geq 0$ is her inconvenience cost per unit distance, known as her ``detour sensitivity''. For simplicity, we denote $d_i(\{i\};r_{\{i\}})$ by $S_iD_i$.

The cost sharing scheme $f$ is such that, for any subset $S\subseteq \mathcal{N}$, $f(i,S;r_S)$ denotes the portion of $\mathcal{OC}(S;r_S)$ allocated to passenger $i\in S$. We set $f(i,S;r_S)=0$ whenever $i\notin S$.


\vspace{0.05in}
\noindent\textbf{Disutility, IR and SIR:} The definitions of disutility, IR, and SIR from equations~\eqref{eq:DU-gen}-\eqref{eq:SIR-gen} carry over in a straightforward manner to the ridesharing scenario, with the dependence on the route emphasized. Thus, $\mathcal{DU}_i(t_j)$ is the disutility of passenger $i$ along a route $r_{\mathcal{N}}(t_j)$ which is identical to $r_{\mathcal{N}}$ up to time $t_j$, but thereafter does not pick up any more passengers, proceeding only to drop off the remaining passengers at their respective destinations. Also, the disutility corresponding to exclusive service is $\mathcal{DU}_i^0 = f(i,\{i\};r_{\{i\}}) = \alpha_{op} S_iD_i$.

\begin{definition}
A route $r_{\mathcal{N}}$ is \textit{IR-feasible} (respectively, \textit{SIR-feasible}) if there exists a budget-balanced cost sharing scheme $f$ that is IR (respectively, SIR) on $r_{\mathcal{N}}$.
\end{definition}

From here on, whenever it is understood from context, we drop the explicit dependence on the route to simplify notation. Next, we present an illustrative example.

\noindent\hrulefill
\begin{example}\label{eg:model-example}
Consider $n=3$ passengers, picked up from their sources $S_1$, $S_2$, $S_3$ (in that order),and travelling to a common destination $D$. The progression of the route $r_{\mathcal{N}}(t)$, as the passengers are picked up one by one, is depicted in Fig.~\ref{fig:model-example}. Given the final route $r_{\mathcal{N}}$, the total distances traveled by passengers $1,2$ and $3$ are $d_1(\mathcal{N}) = S_1S_2 + S_2S_3 + S_3D, d_2(\mathcal{N}) = S_2S_3+S_3D$ and $d_3(\mathcal{N}) = S_3D$. The total distance traveled is $d(\mathcal{N}) = S_1S_2 + S_2S_3 + S_3D$. The operational cost is thus $\mathcal{OC}(\mathcal{N}) = \alpha_{op}(S_1S_2 + S_2S_3 + S_3D)$. Therefore, if $f$ is a budget balanced cost sharing scheme, $f(1,\mathcal{N})+f(2,\mathcal{N})+f(3,\mathcal{N})=\alpha_{op}(S_1S_2 + S_2S_3 + S_3D)$.

\begin{figure}[ht]
\centering
    \includegraphics[width=0.3\columnwidth]{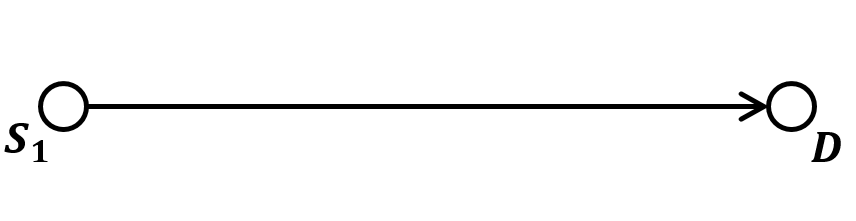}\hfill\includegraphics[width=0.3\columnwidth]{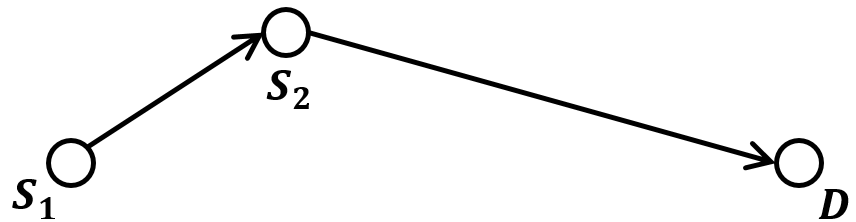}\hfill\includegraphics[width=0.3\columnwidth]{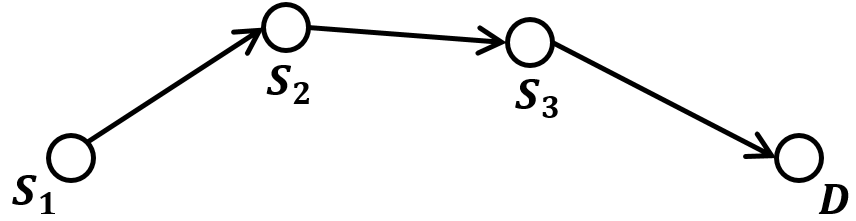}
    \caption{Route progress while picking up passengers traveling to a common destination.\label{fig:model-example}}
\end{figure}

\noindent The inconvenience costs incurred by each passenger due to other passengers are:
\begin{align*}
\mathcal{IC}_{1}(\mathcal{N}) &= \alpha_1(S_1S_2 + S_2S_3 + S_3D - S_1D),\\
\mathcal{IC}_{2}(\mathcal{N}) &= \alpha_2(S_2S_3 + S_3D - S_2D),\\
\mathcal{IC}_{3}(\mathcal{N}) &= \alpha_3(S_3D - S_3D) = 0.
\end{align*}
Thus, a budget-balanced cost sharing scheme $f$ is IR on route $r_{\mathcal{N}}$ if
\begin{align*}
f(1,\mathcal{N}) + \alpha_1(S_1S_2 + S_2S_3 + S_3D - S_1D) \leq \alpha_{op}S_1D,\\
f(2,\mathcal{N}) + \alpha_2(S_2S_3 + S_3D - S_2D) \leq \alpha_{op}S_2D,\\
f(3,\mathcal{N}) \leq \alpha_{op}S_3D.
\end{align*}
The SIR constraints are stronger, since they require IR at every stage of the ride:
\begin{align*}
f(1,\mathcal{N}) + \alpha_1(S_1S_2 + S_2S_3 + S_3D - S_1D) \leq&\ f(1,\mathcal{N}\setminus\{3\}) + \alpha_1(S_1S_2 + S_2D - S_1D) \leq \alpha_{op}S_1D,\\
f(2,\mathcal{N}) + \alpha_2(S_2S_3 + S_3D - S_2D) \leq&\ f(2,\mathcal{N}\setminus\{3\}) \leq \alpha_{op}S_2D,\\
f(3,\mathcal{N}) \leq&\ \alpha_{op}S_3D.
\end{align*}
A necessary condition for the route to be SIR-feasible is therefore obtained by summing up these inequalities (at each stage), using budget-balance of $f$, and simplifying:
\begin{equation*}
S_2S_3+S_3D-S_2D \leq \frac{\alpha_{op}}{\alpha_{op}+\alpha_{1}+\alpha_{2}}S_3D \quad\mbox{and}\quad S_1S_2+S_2D-S_1D \leq \frac{\alpha_{op}}{\alpha_{op}+\alpha_{1}}S_2D.
\end{equation*}
These ``triangle inequalities'' can be interpreted as imposing upper bounds on the \textit{incremental detours} at every stage of the ride. We discuss this in more detail in Section~\ref{SEC:SINGLE}.
\end{example}
\noindent\hrulefill 

\subsection{Characterizing SIR-Feasibile Routes}\label{sec:sir}

The intuition gained from Example~\ref{eg:model-example} suggests that routes with ``large'' detours are unlikely to be SIR-feasible, that is, no budget-balanced cost sharing scheme would be SIR on such routes. Theorem~\ref{THM:SIR-FEASIBILITY} provides a formal characterization of SIR-feasible routes.\footnote{Such a characterization would be useful to augment a routing algorithm in suggesting SIR-feasible routes (when grouping ridesharing requests and assigning them to vehicles).}

\begin{theorem}\label{THM:SIR-FEASIBILITY}
A route $r_{\mathcal{N}}$ is SIR-feasible if and only if
\begin{equation}\label{eq:sir-feasible-constraints}
\begin{split}
\alpha_{op}\Big(d(S(j)) - d(S(j-1))\Big) &+ \sum_{i=1}^{j-1}\alpha_i\Big(d_i(S(j))-d_i(S(j-1))\Big)\\
&\leq \alpha_{op}d(\{j\}) - \alpha_j\Big(d_j(S(j))-d(\{j\})\Big),\quad \forall\ 2\leq j\leq n.
\end{split}
\end{equation}
\end{theorem}

\noindent The proof is from expanding SIR constraints and algebraic manipulation (Appendix \ref{appendix:proof-thm-sir-feasibility}).

Note that the SIR-feasibility equation~\eqref{eq:sir-feasible-constraints} only guarantees that there \textit{exists} a budget balanced cost sharing scheme that is SIR on route $r_{\mathcal{N}}$. To check whether a specific cost sharing scheme is SIR, we would have to go back to the individual constraints~\eqref{eq:SIR-gen}.

The recursive nature of the SIR-feasibility equation~\eqref{eq:sir-feasible-constraints} makes it particularly easy to be incorporated into practical routing algorithms that involve sequential decision making, especially in dynamic ridesharing (see Section~\ref{SEC:ALG-SIR}). 

\section{The Single Dropoff Scenario}\label{SEC:SINGLE}

The previous section illustrates the complexity of the most general case, where the route $r_{\mathcal{N}}$ consists of multiple pickup and dropoff points. Unfortunately, this complexity makes it difficult to infer a useful interpretation of the SIR-feasibility constraints~\eqref{eq:sir-feasible-constraints}. Thus, in this section, we consider the special case where all passengers $1\leq j\leq n$ are traveling to a common destination $D_j=D$, which exposes an interesting property of SIR, namely, that SIR translates to ``natural'' bounds on the incremental detours. We begin this section by simplifying the general expressions introduced in Section~\ref{sec:model-ridesharing} to the single dropoff scenario.

We denote the distance between any two locations $A$ and $B$ in the underlying metric space by $AB$. Recall that $S(j)=\{1,2,\ldots,j\}$, and that we hide the explicit dependence on the routes to simplify notation. Thus, the distance functions become
\begin{equation}\label{eq:dist-single-dropoff}
d(S(j)) = \sum_{k=1}^{j-1}S_kS_{k+1}+S_jD,\quad 1\leq j\leq n,\quad\mbox{and}\quad d_i(S(j)) = \sum_{k=i}^{j-1}S_kS_{k+1}+S_jD,\quad 1\leq i\leq j\leq n.
\end{equation}
The cost functions then become
\begin{equation*}
\begin{split}
\mathcal{OC}(S(j)) &= \alpha_{op}d(S(j)) = \alpha_{op}\left(\sum_{k=1}^{j-1}S_kS_{k+1}+S_jD\right),\quad 1\leq j\leq n.\\
\mathcal{IC}_i(S(j)) &= \alpha_i(d_i(S(j))-d_i(\{i\})) = \alpha_i\left(\sum_{k=i}^{j-1}S_kS_{k+1}+S_jD-S_iD\right),\quad 1\leq i\leq j\leq n.
\end{split}
\end{equation*}
The disutilities are given by
\begin{equation*}
\mathcal{DU}_i(S(j)) = f(i,S(j)) + \alpha_i\left(\sum_{k=i}^{j-1}S_kS_{k+1}+S_jD-S_iD\right),\quad 1\leq i\leq j\leq n.
\end{equation*}
The IR constraints for any budget-balanced cost sharing scheme simplify to
\begin{equation}\label{eqn:IRsd}
f(i,\mathcal{N}) + \alpha_i\left(\sum_{k=i}^{n-1}S_kS_{k+1}+S_nD-S_iD\right) \leq f(i,\{i\}) = \alpha_{op}S_iD,\quad 1\leq i\leq n.
\end{equation}
The SIR constraints~\eqref{eq:sir-expand-1}~and~\eqref{eq:sir-expand-2} for a budget-balanced cost sharing scheme simplify to
\begin{equation*}
\begin{split}
\Big(f(i,S(j)) - f(i,S(j-1))\Big) + \alpha_i\left(S_{j-1}S_j + S_jD - S_{j-1}D\right) \leq 0,\quad &1\leq i < j\leq n.\\
f(j,S(j)) \leq \alpha_{op}S_jD,\quad &2\leq j\leq n.
\end{split}
\end{equation*}
Finally, the SIR-feasibility constraints~\eqref{eq:sir-feasible-constraints} from Theorem~\ref{THM:SIR-FEASIBILITY} simplify to
\begin{equation}\label{eq:sir-feasible-constraints-single}
S_{j-1}S_j + S_jD - S_{j-1}D \leq \frac{S_jD}{1+\frac{1}{\alpha_{op}}\sum_{k=1}^{j-1}\alpha_k},\quad 2\leq j\leq n.
\end{equation}
Notice that when restricted to the single dropoff scenario, the SIR-feasibility constraints assume a much simpler form. For each $j$, the constraint has terms involving only $j$ and $j-1$. This ``Markovian'' nature could prove useful when studying algorithmic problems relating to SIR (see Section~\ref{SEC:ALG-SIR}).

\begin{figure}[ht]
\centering
    \includegraphics[width=0.25\columnwidth]{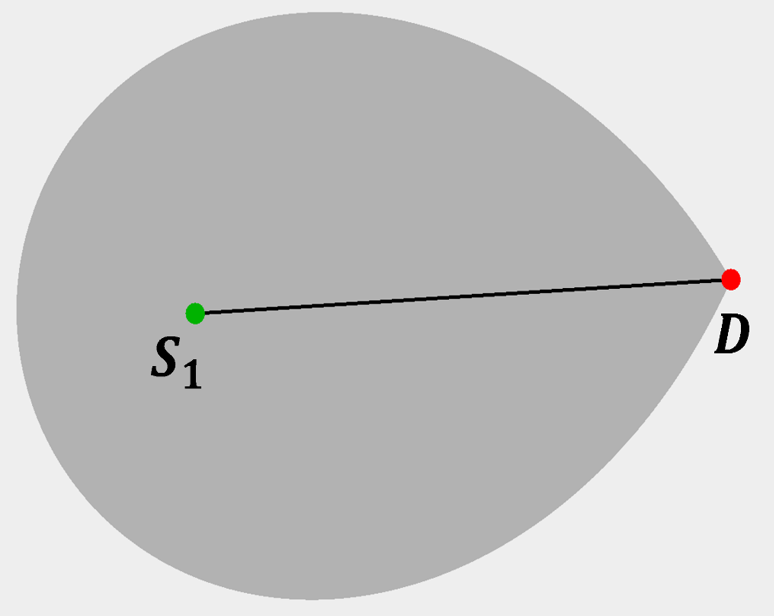}\hspace{0.2in}\includegraphics[width=0.25\columnwidth]{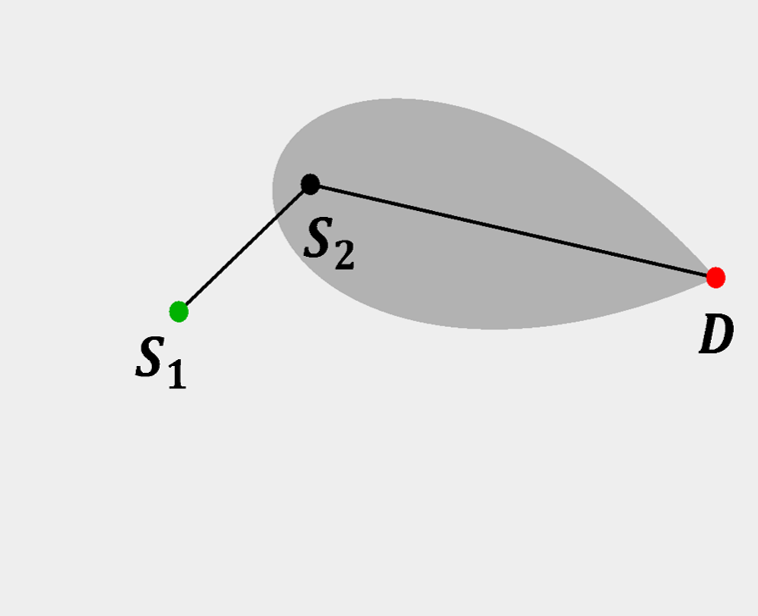}\hspace{0.2in}\includegraphics[width=0.25\columnwidth]{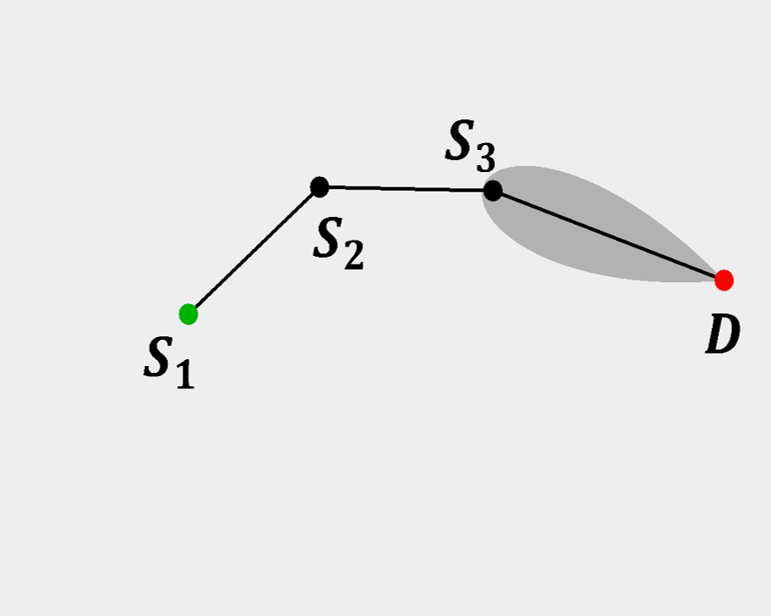}
    \caption{Evolution of the SIR-feasible region (dark shade) while picking up passengers that are traveling to a common destination. Note that the region diminishes rapidly with every subsequent pickup.\label{fig:sir-example}}
\end{figure}

Upon closer inspection, we note that the left hand side of the SIR-feasibility constraints~\eqref{eq:sir-feasible-constraints-single} are nothing but the incremental detours due to picking up subsequent passengers $j$. Thus,~\eqref{eq:sir-feasible-constraints-single} can be viewed as imposing an upper bound on the permissible incremental detour involved in picking up passenger $j$. This bound diminishes with increasing $j$ and increasing proximity to the destination, which means that as more passengers are picked up, the permissible additional detour to pick up yet another passenger keeps shrinking, which is natural. For the passengers in Example~\ref{eg:model-example}, Fig.~\ref{fig:sir-example} shows the evolution of the ``SIR-feasible region'' (points from which the next passenger can be picked up so that the resultant route is SIR-feasible) in Euclidean space, when $\alpha_j=\alpha_{op}$ for $j=1,2,3$. The shape resembles that of a rotated teardrop.

\subsection{Bounds on Total Distance Traveled along SIR-Feasible Routes}\label{sec:starvation-bounds}

The bounds on incremental detours given by the SIR-feasibility constraints~\eqref{eq:sir-feasible-constraints-single} can be combined to obtain bounds on the total distance traveled by a passenger $i\in \mathcal{N}$ along any SIR-feasible route, as a fraction of their direct travel distance $S_iD$. We call this measure the ``starvation factor'' of passenger~$i$. 
The starvation factor of a route is the maximum starvation factor among all the passengers. Intuitively, the starvation factor of a route is a decreasing function of the ratios $\frac{\alpha_k}{\alpha_{op}}$, since the permissible detours are, from~\eqref{eq:sir-feasible-constraints-single}. That is, passengers that are more sensitive to detours should suffer smaller starvation factors. Our goal in this section is to quantify this intuition.

Let $\mathcal{I}(n)$ denote the space of all single dropoff instances of size $n$ (consisting of $n$ pickup points and a common dropoff point from an underlying metric space). Given an instance $p\in\mathcal{I}(n)$, let $\mathcal{R}(p)$ denote the set of all SIR-feasible routes for this instance.

Given an SIR-feasible route $r\in\mathcal{R}(p)$, let $\gamma_r(i)=\frac{d_i(\mathcal{N};r)}{S_iD}$ denote the starvation factor of passenger $i$ along route $r$, where $d_i(\mathcal{N};r) = \sum_{k=i}^{n-1}S_kS_{k+1}+S_nD$, from~\eqref{eq:dist-single-dropoff}, and let $\gamma_r=\max_{i\in\mathcal{N}}\gamma_r(i)$ denote the starvation factor of the route $r$.

\begin{definition}\label{def:SIR-starvation}
The \textit{SIR-starvation factor} over all single dropoff instances of size $n$ is
\begin{equation*}
\gamma(n) = \max_{p\in\mathcal{I}(n)}\min_{r\in\mathcal{R}(p)}\gamma_r.
\end{equation*}
\end{definition}

We show the following bounds for $\gamma(n)$:
\begin{enumerate}
\item \textbf{Upper Bounds:} (Theorems~\ref{thm:SIR-starvation-no-detour}-\ref{thm:SIR-starvation-upper-2}) The worst starvation factor among SIR-feasible routes, $\max_{p\in\mathcal{I}(n)}\max_{r\in\mathcal{R}(p)}\gamma_r$, is (i)~$\Theta(2^n)$ when $\frac{\alpha_i}{\alpha_{op}}\rightarrow 0$, (ii)~$\Theta(\sqrt{n})$ when $\frac{\alpha_i}{\alpha_{op}}=1$, and (iii)~$1$ when $\frac{\alpha_i}{\alpha_{op}}\rightarrow\infty$, for all $i\in\mathcal{N}$. As upper bounds for $\gamma(n)$, these are not necessarily tight, since an instance for which an SIR-feasible route has the worst starvation factor may also admit other SIR-feasible routes with smaller starvation factors.
\item \textbf{Lower Bounds:} (Theorem~\ref{thm:SIR-starvation-lower}) $\gamma(n)$ is no smaller than (i)~$\Theta(n)$ when $\frac{\alpha_i}{\alpha_{op}}\rightarrow 0$, and (ii)~$\Theta(\log n)$ when $\frac{\alpha_i}{\alpha_{op}}=1$, for all $i\in\mathcal{N}$. These lower bounds are tight.
\end{enumerate}
It is interesting to note that the gap between the upper and lower bounds narrows down and vanishes as $\frac{\alpha_i}{\alpha_{op}}$ increases to $\infty$.\footnote{Note that, by definition, $1$ is always a trivial lower bound for the starvation factor of any route, since the points are from an underlying metric space.} The proofs are complex, and can be found in Appendix~\ref{app:proofs-sec-4}.

We begin by establishing an almost obvious result that when passengers are infinitely inconvenienced by even the smallest of detours,\footnote{Frankly, why would such passengers even consider ridesharing?} the only SIR-feasible routes (indeed, even IR-feasible routes) are those with zero detours, which implies a starvation factor of $1$.

\begin{theorem}\label{thm:SIR-starvation-no-detour}
If $\frac{\alpha_i}{\alpha_{op}}\rightarrow\infty$ for all $i\in\mathcal{N}$, then $\gamma_r=1$ for any SIR-feasible route $r$.
\end{theorem}

Next, we consider passengers who value their time more than $\alpha_{op}$, and show that the worst they would have to endure is a \textit{sublinear} starvation factor, in particular, $\Theta(\sqrt{n})$. This is tight, that is, there exists an SIR-feasible route with $\Theta(\sqrt{n})$ starvation factor, when $\alpha_i=\alpha_{op}$ for all $i\in\mathcal{N}$. However, as the $\alpha_i$ keep increasing beyond $\alpha_{op}$, this bound becomes looser, culminating in a $\Theta(\sqrt{n})$ gap when $\alpha_i\rightarrow\infty$, as evidenced by Theorem~\ref{thm:SIR-starvation-no-detour}.

\begin{theorem}\label{thm:SIR-starvation-upper-1}
If $\frac{\alpha_i}{\alpha_{op}}\geq 1$ for all $i\in\mathcal{N}$, then $\gamma_r \leq 2\sqrt{n}$ for any SIR-feasible route $r$.
\end{theorem}

Even though it may be unrealistic, as an academic exercise, we investigate an upper bound on $\gamma_r$ when the passengers are completely unaffected by detours, that is, $\frac{\alpha_i}{\alpha_{op}}\rightarrow 0$ for all $i\in\mathcal{N}$. Not surprisingly, it turns out that the starvation factor can be exponentially large in such a scenario, as the next theorem shows.

\begin{theorem}\label{thm:SIR-starvation-upper-2}
If $\frac{\alpha_i}{\alpha_{op}}\rightarrow 0$ for all $i\in\mathcal{N}$, then $\gamma_r \leq 2^{n}$ for any SIR-feasible route $r$.
\end{theorem}

\begin{figure}[ht]
\centering
    \includegraphics[width=0.65\columnwidth]{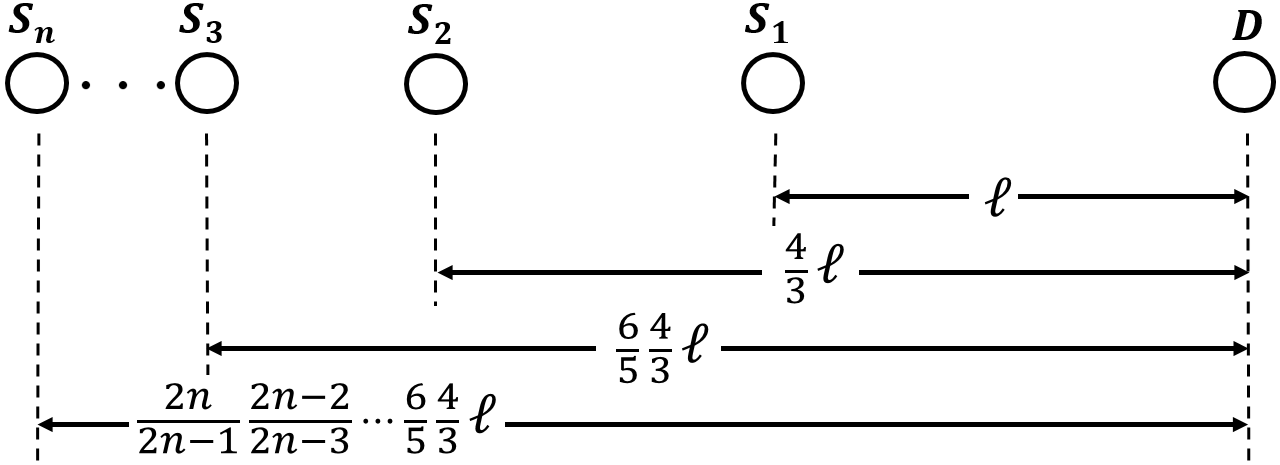}
        \caption{Single dropoff instance with a route $(S_1,S_2,\ldots,S_n,D)$ whose starvation factor is $\Theta(\sqrt{n})$. If the distances $S_iD$, $1\leq i\leq n$, were $2^{i-1}\ell$ instead, then the starvation factor of the same route would be $\Theta(2^{n})$.\label{fig:SIR-starvation-upper}}
\end{figure}

The upper bounds of Theorems~\ref{thm:SIR-starvation-upper-1}-\ref{thm:SIR-starvation-upper-2} on $\gamma_r$ are tight, as discussed next; however, by Definition~\ref{def:SIR-starvation}, they also serve as upper bounds on $\gamma(n)$, in which capacity, they may not necessarily be tight. This is because, an instance for which an SIR-feasible route has the worst starvation factor may also admit better SIR-feasible routes. For example, Fig.~\ref{fig:SIR-starvation-upper} depicts an instance in one-dimensional Euclidean space for which the route $(S_1,S_2,\ldots,S_n,D)$ is SIR-feasible (satisfying~\eqref{eq:sir-feasible-constraints-single-eq} with equality) and has a starvation factor of $\Theta(\sqrt{n})$. (The same instance with the distances appropriately modified illustrates the $\Theta(2^n)$ starvation factor of Theorem~\ref{thm:SIR-starvation-upper-2}.) However, note that the reverse route $(S_n,S_{n-1},\ldots,S_1,D)$ is also SIR-feasible and has a starvation factor of $1$.

Finally, we establish a tight lower bound on $\gamma(n)$ for arbitrary $\alpha_i>0$, by exhibiting an instance with a unique SIR-feasible route with the desired starvation factor.

\begin{theorem}\label{thm:SIR-starvation-lower}
$\gamma(n) \geq \sum_{j=1}^{n}\left(1+\frac{1}{\alpha_{op}}\sum_{k=1}^{j-1}\alpha_k\right)^{-1}$.
\end{theorem}

It is easy to observe that the lower bound of Theorem~\ref{thm:SIR-starvation-lower} simplifies to $\Theta(\log n)$ when $\frac{\alpha_i}{\alpha_{op}}=1$, and $\Theta(n)$ when $\frac{\alpha_i}{\alpha_{op}}\rightarrow 0$, for all $i\in\mathcal{N}$.

\section{The Benefit of Ridesharing and Sequential Fairness}\label{sec:sequential-fairness}

In this section, we explore the consequences of sequential fairness (defined in Section~\ref{sec:sequential-fairness-gen}) on the design of cost sharing schemes for ridesharing, in the single dropoff scenario. First, for $2\leq j\leq n$, $1\leq i\leq j$, the expression for incremental benefit to passenger $i$ due to the addition of passenger $j$ is
\begin{equation}\label{eq:ib}
\mathcal{IB}_{i}(S(j)) =
\begin{cases}
f(i,S(j-1)) - f(i,S(j)) - \alpha_i\left(S_{j-1}S_j + S_jD - S_{j-1}D\right), & 1\leq i < j\\
\alpha_{op}S_jD - f(j,S(j)), & i=j.
\end{cases}
\end{equation}
Thus, for $2\leq j\leq n$, the total incremental benefit due to the addition of passenger $j$ is given by
\begin{equation}\label{eq:tib}
\mathcal{TIB}(S(j)) = \sum_{k=1}^{j} \mathcal{IB}_k(S(j)) = \alpha_{op}S_jD - \left(\alpha_{op}+\sum_{k=1}^{j-1}\alpha_k\right)\left(S_{j-1}S_j + S_jD - S_{j-1}D\right),
\end{equation}
where the dependence on $f$ vanishes due to budget-balance.

For the single dropoff scenario, the incremental inconvenience cost to $i$ due to the detour caused by $j$ is given by $\mathcal{IC}_i(S_{ij})-\mathcal{IC}_i(S_{i(j-1)}) = \alpha_i\left(S_{j-1}S_j+S_jD-S_{j-1}D\right)$; hence, Definition~\ref{def:SF-gen} simplifies to the following equivalent definition.

\begin{definition}\label{def:sequential-fairness}
Given a vector $\vec{\beta}=\left(\beta_2,\beta_3,\ldots,\beta_n\right)$, where $0\leq\beta_j\leq 1$ for $2\leq j\leq n$, a budget balanced cost sharing scheme $f$ is \textit{$\vec{\beta}$-sequentially fair} if, on any SIR-feasible route,
\begin{equation*}
\left(\forall\ 2\leq j\leq n\right)\quad \frac{\mathcal{IB}_{i}(S(j))}{\mathcal{TIB}(S(j))} =
\begin{cases}
\beta_j\frac{\alpha_i}{\sum_{m=1}^{j-1}\alpha_m}, & 1\leq i\leq j-1\\
1-\beta_j, & i=j.
\end{cases}
\end{equation*}
\end{definition}

Note that $1-\beta_j$ denotes the fraction of the total incremental benefit enjoyed by the new passenger $j$ as a result of having them join the ride, and $\beta_j$ denotes the remaining fraction, which is split among the existing passengers in proportion to their $\alpha_i$ values.

It turns out that the requirements imposed by Definition~\ref{def:sequential-fairness}, while perhaps appearing to be quite lenient, are sufficient for a strong and meaningful characterization of sequentially fair cost sharing schemes, as we discuss next.

\subsection{Characterizing Sequentially Fair Cost Sharing Schemes}

We begin this section with a theorem that provides an exact characterization of budget balanced sequentially fair cost sharing schemes for single dropoff scenarios.

\begin{theorem}\label{thm:sf-char}
Given a vector $\vec{\beta}=\left(\beta_2,\beta_3,\ldots,\beta_n\right)$, where $0\leq\beta_j\leq 1$ for $2\leq j\leq n$, a budget-balanced cost sharing scheme $f$ is \textit{$\vec{\beta}$-sequentially fair} if and only if, for $2\leq j\leq n$,
\begin{itemize}
\item The cost to the incoming passenger $j$ is given by
    \begin{small}
    \begin{equation}\label{eq:sf-initialcost}
    f(j,S(j)) = \beta_j\big[\alpha_{op}S_jD\big] + (1-\beta_j)\left[\left(\alpha_{op}+\sum_{m=1}^{j-1}\alpha_m\right)\left(S_{j-1}S_j+S_jD-S_{j-1}D\right)\right].
    \end{equation}
    \end{small}
\item The incremental ``discount'' to each existing passenger $1\leq i\leq j-1$ is given by
    \begin{small}
    \begin{equation}\label{eq:sf-discount}
    \begin{split}
    f(i,S(j-1)) - f(i,S(j))\ =\ &\beta_j\left[\frac{\alpha_i}{\sum_{m=1}^{j-1}\alpha_m}\left(\alpha_{op}S_jD-\alpha_{op}(S_{j-1}S_j+S_jD-S_{j-1}D)\right)\right]\\
    &\qquad\qquad+ (1-\beta_j)\big[\alpha_i\left(S_{j-1}S_j+S_jD-S_{j-1}D\right)\big].
    \end{split}
    \end{equation}
    \end{small}
\end{itemize}
\end{theorem}

We omit the proof, since it is simply a straightforward substitution of equations~\eqref{eq:ib}-\eqref{eq:tib} in Definition~\ref{def:sequential-fairness} and rearrangement of the terms. The characterization of Theorem~\ref{thm:sf-char} reveals elegant structural properties of sequentially fair cost sharing schemes:
\begin{enumerate}[(a)]
\item \textbf{Online Implementability.} When a passenger $j$ is picked up, their estimated cost is given by $f(j,S(j))$, which is their final payment if there are no more pickups. At the same time, each existing passenger $i$ is offered a ``discount'' in the amount of $f(i,S(j-1)) - f(i,S(j))$ that brings down their earlier cost estimates. This suggests a novel ``reverse-meter'' design for a ridesharing application on each passenger's smartphone that keeps track of their estimated final payment, as the ride progresses. Starting with $f(i,S(i))$ when passenger $i$ begins their ride, it would keep decreasing every time a detour begins to pick up a new passenger. Such a visually compelling interface would encourage wider adoption of ridesharing.
\item \textbf{Convex Combination of Extreme Schemes.} For each $j$, $2\leq j\leq n$, the cost sharing scheme is a convex combination of the following two extreme schemes:
    \begin{itemize}
    \item \textit{The total incremental benefit is fully enjoyed by the incoming passenger $j$, i.e., $\beta_j=0$.} Here, from~\eqref{eq:sf-initialcost}-\eqref{eq:sf-discount}, the incoming passenger $j$ (a)~pays the service provider an amount $\alpha_{op}(S_{j-1}S_j+S_jD-S_{j-1}D)$ that corresponds to the increase in the operational cost, and (b)~pays each existing passenger $1\leq i\leq j-1$ an amount $\alpha_{i}(S_{j-1}S_j+S_jD-S_{j-1}D)$ that corresponds to the incremental inconvenience cost they suffered.
    \item \textit{The total incremental benefit is fully enjoyed by the existing passengers $1\leq i\leq j-1$, i.e., $\beta_j=1$.} Here, from~\eqref{eq:sf-initialcost}-\eqref{eq:sf-discount}, the incoming passenger $j$ pays $\alpha_{op}S_jD$, the same as they would have paid for a private ride. From this, the service provider recovers $\alpha_{op}(S_{j-1}S_j+S_jD-S_{j-1}D)$ that corresponds to the increase in the operational cost, and what is left is split among the existing passengers proportional to their $\alpha_i$ values.
    \end{itemize}
    Note that the incoming passenger $j$ pays the least in the former scheme ($\beta_j=0$) and the most in the latter scheme ($\beta_j=1$).
\item \textbf{Transfers Between Passengers.} From the previous observation, it follows that incoming passengers must, at minimum, fully compensate existing passengers for the incremental inconvenience costs that resulted from the detour to pick them up, which can be viewed as internal transfers between passengers. Even though it may be reasonable to expect this from a fair cost sharing scheme, it is remarkable that sequential fairness \textit{mandates} this property.
\end{enumerate}

In designing a sequentially fair cost sharing scheme, $\vec{\beta}$ can be chosen strategically to incentivize commuters to rideshare. A commonly used incentive is to guarantee a minimum discount on the cost of a private ride. In our framework of sequentially fair cost sharing schemes, it corresponds to setting $\beta_j$ so that $f(j,S(j))$ is a desired fraction of $\alpha_{op}S_jD$.\footnote{The SIR-feasibility constraints would have to be appropriately tightened to guarantee such a discount.} We end this section with an example.

\noindent\hrulefill
\begin{example}\label{EG:XHARE-COST}
Consider the single dropoff scenario, where we also assume that $\alpha_i=\alpha_{op}=1$ for all $i\in\mathcal{N}$. For $1\leq i\leq j\leq n$, we define the cost sharing scheme $f^{\texttt{XC}}$ as:
\begin{small}
\begin{equation*}
f^{\texttt{XC}}(i,S(j)) = \left(\sum_{k=i+1}^{j}\frac{S_{k-1}S_{k}}{k-1} + \frac{S_{j}D}{j}\right) + (i-1)\left(S_{i-1}S_i+S_iD-S_{i-1}D\right)- \left(\sum_{k=i+1}^{j}\left(S_{k-1}S_{k} + S_{k}D - S_{k-1}D\right)\right),
\end{equation*}
\end{small}

The first terms correspond to dividing the operational cost of each segment equally among the ridesharing passengers traveling along that segment. The second terms correspond to the passenger $i$ compensating each of the $i-1$ passengers that were picked up earlier, for the incremental detour they suffered. The last terms correspond to the net compensation received by passenger $i$ from all passengers that were picked up later, for the incremental detours that $i$ suffered.

Intuitively, $f^{\texttt{XC}}$ is a ``fair'' cost sharing scheme. In fact, it can be shown that for $\vec{\beta}=\left(\frac{1}{2},\frac{1}{3},\ldots,\frac{1}{n}\right)$, it is a $\vec{\beta}$-sequentially fair cost sharing scheme:

From~\eqref{eq:sf-initialcost}-\eqref{eq:sf-discount}, we get
\begin{small}
\begin{equation*}
\begin{split}
\frac{\mathcal{IB}_j(S(j))}{\mathcal{TIB}(S(j))} &= \frac{S_jD-f^{\texttt{XC}}(j,S(j))}{S_jD - j\left(S_{j-1}S_j + S_jD - S_{j-1}D\right)}\\
&= \frac{S_jD-\left(\frac{S_{j}D}{j}+(j-1)\left(S_{j-1}S_i+S_jD-S_{j-1}D\right)\right)}{{S_jD - j\left(S_{j-1}S_j + S_jD - S_{j-1}D\right)}} = \frac{j-1}{j} = 1-\frac{1}{j},
\end{split}
\end{equation*}
\end{small}

\noindent as desired. Also, for $1\leq i\leq j-1$, we get
\begin{small}
\begin{equation*}
\begin{split}
\frac{\mathcal{IB}_i(S(j))}{\mathcal{TIB}(S(j))} &= \frac{f^{\texttt{XC}}(i,S(j-1)) - f^{\texttt{XC}}(i,S(j)) - \left(S_{j-1}S_j + S_jD - S_{j-1}D\right)}{S_jD - j\left(S_{j-1}S_j + S_jD - S_{j-1}D\right)}\\
&= \frac{\frac{S_{j-1}D}{j-1}-\left(\frac{S_{j-1}S_j}{j-1}+\frac{S_{j}D}{j}\right)+\left(S_{j-1}S_j + S_jD - S_{j-1}D\right) - \left(S_{j-1}S_j + S_jD - S_{j-1}D\right)}{S_jD - j\left(S_{j-1}S_j + S_jD - S_{j-1}D\right)}\\
&= \frac{\frac{S_{j}D}{j(j-1)}-\frac{1}{j-1}\left(S_{j-1}S_j + S_jD - S_{j-1}D\right)}{S_jD - j\left(S_{j-1}S_j + S_jD - S_{j-1}D\right)} = \frac{1}{j}\frac{1}{j-1}.
\end{split}
\end{equation*}
\end{small}
\end{example}
\noindent\hrulefill 

\section{New Algorithmic Problems}\label{SEC:ALG-SIR}

As discussed in Section~\ref{sec:sir}, the SIR-feasibility constraints~\eqref{eq:sir-feasible-constraints} or~\eqref{eq:sir-feasible-constraints-single}, can be considered as additional constraints to the routing optimization problem. For instance, vehicle routing problems with various operational objectives, ridesharing with multiple pickups and dropoff points, online routing problems can all benefit from incorporating SIR-feasibility constraints while performing route optimization. As a concrete example, consider the following single dropoff ride matching and routing problem:

\textit{Given $n$ pickup points and a common dropoff point in a metric space, (a)~does there exist an allocation of pickup points to $1\leq m\leq n$ vehicles, each with capacity $\lceil\frac{n}{m}\rceil\leq c\leq n$, such that there exists an SIR-feasible route for each vehicle? And (b)~if so, what is the allocation and corresponding routes that minimize the total ``vehicle-miles'' traveled?}

We do not know whether the feasibility problem~(a) can be solved in polynomial time, even when $m=1$ and $\alpha_i=\alpha_j$ for all $1\leq i,j\leq n$, where it reduces to finding a sequence of the pickup points that satisfies the inequalities~\eqref{eq:sir-feasible-constraints-single}. The ``Markovian'' nature of these inequalities (each inequality only depends on adjacent pickup points in the route) suggests that it may be worth trying to come up with a polynomial time algorithm for the feasibility problem. In Section~\ref{ssec:sir-np-hard}, we show that this problem is NP-hard when not restricted to a metric space, which implies that any poly-time algorithm, if one exists, must necessarily exploit the properties of a metric space. However, even if one succeeds in this endeavor, we show in Section~\ref{ssec:opt-sir-np-hard} that the optimization~(b) over all SIR-feasible routes is NP-hard.

Like SIR-feasibility, there might be other constraints on the ordering of the pickup points (for instance, due to hard requirements on pickup times). Studying such variants might help understand how to tackle SIR-feasibility constraints. For example, it is known that finding the optimal allocation (minimizing the total vehicle-miles traveled) of passengers to vehicles without any restriction on the order of pickups is NP-hard~\cite{Cordeau2007}. On the other hand, as we show in Section~\ref{ssec:opt-total-order-allocation}, the problem is polynomial time solvable if a strict total ordering is imposed and the capacity of each vehicle is unrestricted. It then becomes an interesting future direction to investigate what kinds of order constraints retain polynomial time solvability of the problem.

\subsection{Determining Existence of SIR-Feasible Routes is Hard}\label{ssec:sir-np-hard}

In this section, we present Theorem~\ref{thm:sir-feasibility-hard}, which shows that, for the single-dropoff scenario, determining whether an SIR-feasible route exists is NP-hard in general, by a reduction from the undirected Hamiltonian path problem.\footnote{Given an undirected graph, a Hamiltonian path is a path in the graph that visits each vertex exactly once. The undirected Hamiltonian path problem is to determine, given an undirected graph, whether a Hamiltonian path exists. It is known to be NP-hard.} The proof is deferred to Appendix~\ref{app:proofs-sec-6}.

\begin{definition}
Given a set $\mathcal{N}$ of $n$ pickup points, and a common dropoff point in an underlying (possibly non-metric) space, and positive coefficients $\alpha_{op}, \alpha_{1}, \alpha_{2}, \ldots, \alpha_{n}$,  \texttt{SIR-Feasibility} is the problem of determining whether an SIR-feasible route of length $n$ exists, that is, whether there exists a sequence of the pickup points that satisfies the SIR-feasibility constraints~\eqref{eq:sir-feasible-constraints-single}.
\end{definition}

\begin{theorem}\label{thm:sir-feasibility-hard}
\texttt{SIR-Feasibility} is NP-hard.
\end{theorem}

However, it can be easily seen that \texttt{SIR-Feasibility} is not hard in certain special cases and in certain metric spaces. Consider an input graph, where the pickup points and the dropoff point are embedded on a line, and $\alpha_i=\alpha_{op}$ for all $i \in \mathcal{N}$. Without loss of generality, we assume that the pickup points $\{S_1,\ldots,S_n\}$ appear in the same order on the line, so that $S_1$ and $S_n$ are the two end points. Clearly, if the destination $D$ occurs before $S_1$ (respectively, after $S_n$), the instance is SIR-feasible. This is because the route starting from $S_n$ (respectively, $S_1$) and ending at $D$, visiting all the pickup points along the way incurs zero detour for everyone, and is thus SIR-feasible. In fact, such a route also traverses the minimum distance among all feasible routes. However, consider the case where $D$ is located at some intermediate location. Such an instance will never be SIR-feasible. To see this, first consider an instance where $n=2$, and $S_1 < D < S_2$. Let $S_1D = x$, $S_2D = y$; hence $S_1S_2 = x+y$. We analyze the SIR-feasiblity constraints~\eqref{eq:sir-feasible-constraints-single} for each of two cases. If $S_1$ is visited before $S_2$, then SIR-feasibility requires that $x+y + y - x \leq \frac{y}{2}$, which is impossible. Similarly, if $S_2$ is visited before $S_1$, then SIR-feasibility requires that $x+y + x - y \leq \frac{x}{2}$, which is also impossible. Now, when $n>2$ and $D$ is located at an intermediate point, any feasible route must, at some point, ``jump over'' $D$ from some $S_i$ to another $S_j$, at which stage the analysis would be the same as that for $n=2$, and is therefore not SIR-feasible. A similar phenomenon can be observed when the underlying metric is a tree rooted at $D$ and the pickup points are located at the leaves, and $\alpha_i=\alpha_{op}$ for all $i \in \mathcal{N}$. It can be shown that instances where the pickup points are spread across more than one subtree rooted at $D$ cannot be SIR-feasible, and when the pickup points are all part of a single subtree rooted at $D$, SIR-feasibility can be checked in polynomial time. We leave open the problem of determining whether \texttt{SIR-Feasibility} is hard in general metric spaces.

%


\subsection{Optimizing over SIR-Feasible Routes is Hard}\label{ssec:opt-sir-np-hard}

Given an undirected weighted graph, the problem of determining an optimal Hamiltonian cycle\footnote{A Hamiltonian cycle is a Hamiltonian path that is a cycle. In other words, it is a cycle in the graph that visits each vertex exactly once.} (one that minimizes the sum of the weights of its edges) is a well known problem called the \textit{Traveling Salesperson Problem}, abbreviated as \textit{TSP}. A slight variant of this problem, known as \textit{Path-TSP}, is when the traveling salesperson is not necessarily required to return to the starting point or depot, in which case we only seek an optimal Hamiltonian path. These problems are NP-hard~\citep{Papadimitriou1994}. Special cases of the above problems arise when the graph is complete and the edge weights correspond to distances between vertices from a metric space. These variants, which we call \textit{Metric-TSP} and \textit{Metric-Path-TSP}, respectively, are also NP-hard, e.g.,~\cite{Papadimitriou1977} showed the hardness for the Euclidean metric.


\begin{definition}
Given a set $\mathcal{N}$ of $n$ pickup points, a common dropoff point in an underlying metric space, and positive coefficients $\alpha_{op}, \alpha_{1}, \alpha_{2}, \ldots, \alpha_{n}$, \texttt{Opt-SIR-Route} is the problem of finding an SIR-feasible route of length $n$ of minimum total distance.
\end{definition}

\begin{theorem}\label{thm:opt-sir-route-hard}
\texttt{Opt-SIR-Route} is NP-hard.
\end{theorem}

The proof is via a reduction from \textit{Metric-Path-TSP}; see Appendix~\ref{app:proofs-sec-6}.

\subsection{Optimal Allocation of Totally Ordered Passengers to Uncapacitated Vehicles}\label{ssec:opt-total-order-allocation}

In this section, we present a polynomial time algorithm for optimal allocation of passengers to vehicles (minimizing the total vehicle-miles traveled), given a total order on the pickups, and when the capacity of any vehicle is unrestricted. To the best of our knowledge, this result is new; see~\cite{prins2014order} for a survey on related problem variants.

\begin{figure}[ht]
\centering
    \includegraphics[width=0.85\textwidth]{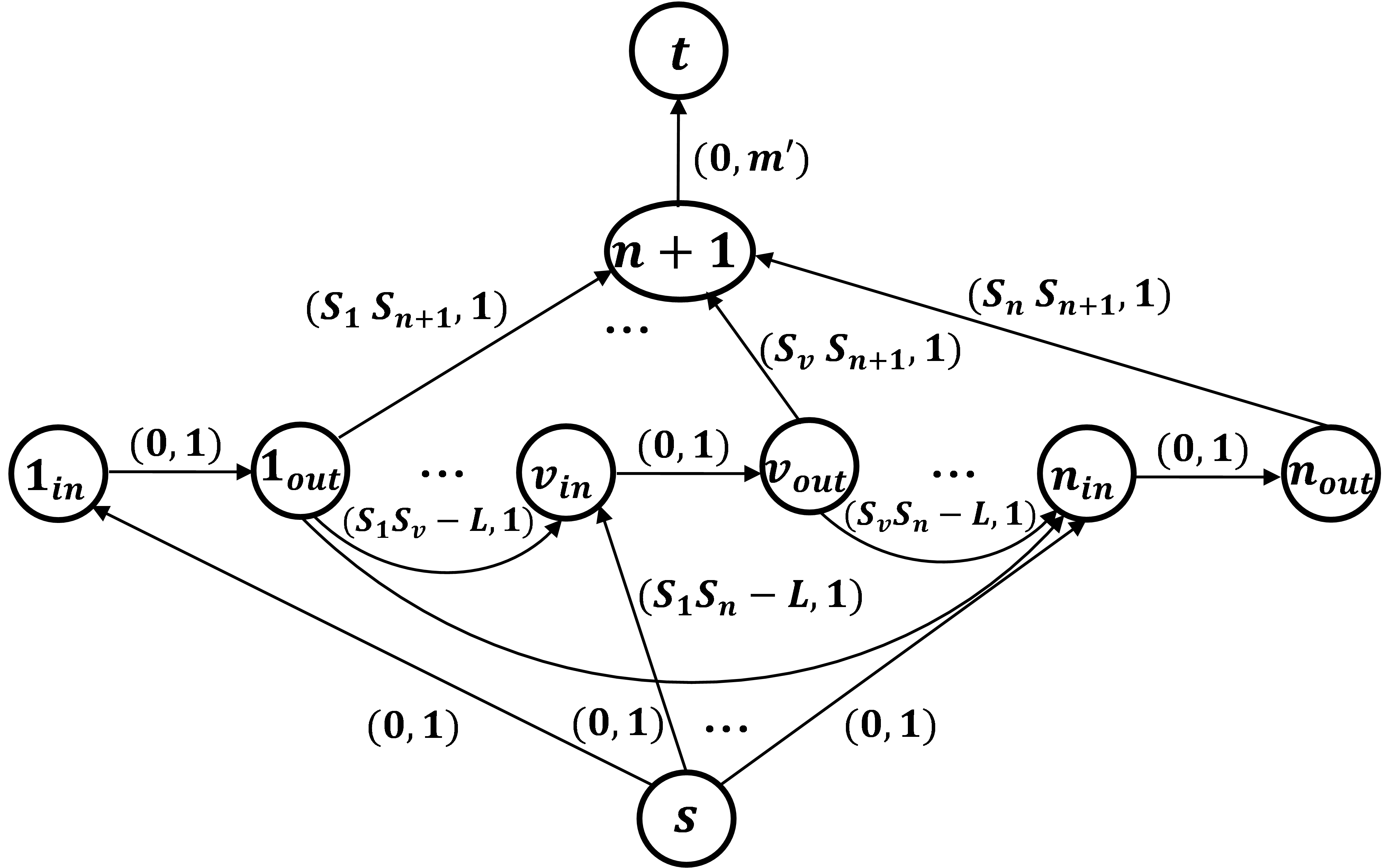}
    \caption{Illustration of the directed acyclic flow network, a minimum cost maximum flow on which corresponds to an assignment of $n$ totally ordered passengers to $m'$ uncapacitated vehicles. Each of the edge labels correspond to a tuple consisting of edge cost and edge capacity.\label{fig:passenger-assignment}}
\end{figure}

Our result relies on reducing the allocation problem to a minimum cost flow problem on a flow network with integral capacities. We are given the set $\mathcal{N}$ of passengers (that is, the set of $n$ ordered pickup locations) traveling to a single dropoff location $D$. Without loss of generality, we let the indices in $\mathcal{N}$ reflect the position in the pickup order, that is, $u\in\mathcal{N}$ is the $u$-th pick up from location $S_u$. For convenience, we index the destination $D$ as $n+1$. Let the unknown optimal assignment use $1\leq m' \leq n$ vehicles (we address how to find it later). A directed acyclic flow network (see Figure~\ref{fig:passenger-assignment}) is then constructed as follows:
\begin{enumerate}[(1)]
\item $s$ and $t$ denote the source and sink vertices, respectively.
\item For each passenger/pickup location $u\in\mathcal{N}$, we create two vertices and an edge: an entry vertex $u_{in}$, an exit vertex $u_{out}$, and an edge of cost $0$ and capacity $1$ directed from $u_{in}$ to $u_{out}$. We also create a vertex $n+1$ corresponding to the dropoff location.
\item We create $n$ edges, one each of cost $0$ and capacity $1$ from the source vertex $s$ to each of the entry vertices $u_{in}$, $u\in\mathcal{N}$.
\item We create $n$ edges, one each of cost $S_uD$ and capacity $1$ from each of the exit vertices $u_{out}$, $u\in\mathcal{N}$, to the dropoff vertex $n+1$.
\item To encode the pickup order, for each $1\leq u < v\leq n$ we create an edge of cost $(S_uS_v-L)$ and capacity $1$ directed from $u_{out}$ to $v_{in}$, where $L$ is a sufficiently large number satisfying $L>2\max_{u,v \in \mathcal{N}\cup{\{n+1\}}}S_uS_v$.
\item We add a final edge of cost $0$ and capacity $m'$ from the dropoff vertex $n+1$ to the sink vertex $t$, thereby limiting the maximum flow in the network to $m'$ units.
\end{enumerate}
Since all the edge capacities are integral, the integrality theorem guarantees an integral minimum cost maximum flow, and we assume access to a poly-time algorithm to compute it in a network with possibly negative costs on edges. Notice that we do have negative edge costs (step~(5) of the above construction); however, our network is a directed acyclic graph, owing to the fact that there is a total ordering on the pickup locations. Hence, there are no negative cost cycles.

We defer the full proof to Appendix~\ref{appendix:proof-reduction}; however, we briefly outline the steps involved:
\begin{itemize}
\item Any integral maximum flow from $s$ to $t$ must be comprised of $m'$ vertex-disjoint paths between the source vertex $s$ and the dropoff vertex $n+1$.
\item Any integral minimum cost flow must cover all the $2n$ pickup vertices, that is, a unit of flow enters every entry vertex $u_{in}$, and a unit of flow exits each exit vertex $u_{out}$, $u\in\mathcal{N}$.
\item The partition of $\mathcal{N}$ according to the $m'$ vertex-disjoint paths between $s$ and $n+1$ in an integral minimum cost maximum flow corresponds to the optimal allocation of the $n$ totally ordered passengers among $m'$ uncapacitated vehicles.
\end{itemize}
Finally, we argue that the overall optimal assignment can be obtained by computing the optimal assignments using the above reduction for each $1\leq m'\leq n$ and choosing the one with the overall minimum cost, which completes the reduction.

\section{Concluding Remarks}\label{sec:conclusion}

In addition to the discussion in Section~\ref{ssec:weaker-SIR-SF-gen}, and the open algorithmic questions raised in the previous section, there are a few other important aspects that we believe future work should address. We conclude the paper with brief discussions of these issues.

Throughout, we have assumed knowledge of the passengers' $\alpha_i$ values; but in reality, they are most likely private information, especially in commercial ridesharing. One can either attempt to learn these values over time from passenger feedback, or, one can ask the passenger for this information. In the latter case, truthful reporting is a concern. Since our framework explicitly takes into account the inconvenience costs of passengers while considering SIR-feasibility as well as sequential fairness, no sequentially fair cost sharing scheme can be dominant strategy incentive compatible. It would be interesting to study the trade-offs between efficiency, fairness, budget-balance, and incentive compatibility. The best framework in which to study these questions (including in more general settings where there is uncertainty about future pickup requests) is perhaps online mechanism design~\citep{Parkes2007,dzhao15,shen16}.

The duality between cost sharing and benefit sharing in our framework is worth a deeper analysis. While the space of cost sharing schemes that the two views accommodate are no different from each other, there is a crucial difference in approaching their design. In particular, notice that a budget balanced cost sharing scheme need only recover the operational costs; see~\eqref{eq:BB-gen}. The inconvenience costs experienced by the passengers are a separate artifact of our framework, which only explicitly affect the design of cost sharing schemes when viewed through the lens of benefit sharing and sequential fairness. What ``traditional'' fairness properties does a sequentially fair cost sharing scheme possess? Under what conditions, if any, is it a (generalized) Shapley value, or is in the core?

\begin{APPENDICES}

\section{Proofs from Section~\ref{SEC:SINGLE}}\label{app:proofs-sec-4}

\subsection{Proof of Theorem \ref{THM:SIR-FEASIBILITY}}
\label{appendix:proof-thm-sir-feasibility}

The proof follows from expanding the SIR constraints~\eqref{eq:SIR-gen}. First, for any $2\leq j\leq n$, and $1\leq i\leq j-1$, the SIR constraint can be expanded as
\begin{equation}\label{eq:sir-expand-1}
\begin{split}
& f(i,S(j)) + \mathcal{IC}_i(S(j)) \leq f(i,S(j-1)) + \mathcal{IC}_i(S(j-1))\\
\Longrightarrow\;\; & \Big(f(i,S(j))-f(i,S(j-1))\Big) + \Big(\mathcal{IC}_i(S(j)) - \mathcal{IC}_i(S(j-1))\Big) \leq 0.
\end{split}
\end{equation}
For $i=j$, the SIR constraint can be expanded as
\begin{equation}\label{eq:sir-expand-2}
\begin{split}
& f(j,S(j)) + \mathcal{IC}_j(S(j)) \leq f(j,\{j\})\\
\Longrightarrow\;\; & f(j,S(j)) \leq \alpha_{op}d(\{j\}) - \mathcal{IC}_j(S(j)),
\end{split}
\end{equation}
where, it follows from budget-balance that $f(j,\{j\})=\alpha_{op}d(\{j\})$.

The ``only if'' direction can be seen to hold by adding all the $j$ inequalities given by~\eqref{eq:sir-expand-1}~and~\eqref{eq:sir-expand-2}, for all $2\leq j\leq n$.
%
\begin{equation*}
\Big(\sum_{i=1}^{j}f(i,S(j)) - \sum_{i=1}^{j-1}f(i,S(j-1))\Big) + \sum_{i=1}^{j-1}\Big(\mathcal{IC}_i(S(j)) - \mathcal{IC}_i(S(j-1))\Big) \leq \alpha_{op}d(\{j\}) - \mathcal{IC}_j(S(j)).
\end{equation*}
Using the budget-balance property~\eqref{eq:BB-gen} to simplify the first two terms, we get
\begin{equation}\label{eq:sir-feasible-constraints-alt}
\Big(\mathcal{OC}(S(j))-\mathcal{OC}(S(j-1))\Big) + \sum_{i=1}^{j-1}\Big(\mathcal{IC}_i(S(j)) - \mathcal{IC}_i(S(j-1))\Big) \leq \alpha_{op}d(\{j\}) - \mathcal{IC}_j(S(j)).
\end{equation}
Equation~\eqref{eq:sir-feasible-constraints} then follows by substituting for $\mathcal{OC}(\cdot)$ and $\mathcal{IC}_i(\cdot)$ from~\eqref{eq:oc}~and~\eqref{eq:ic} respectively, and simplifying.

Next, we prove the ``if'' direction. Assuming that~\eqref{eq:sir-feasible-constraints} holds, or, alternatively, assuming that~\eqref{eq:sir-feasible-constraints-alt} holds, it suffices to exhibit a budget-balanced cost sharing scheme $f$, under which all the SIR constraints given by~\eqref{eq:sir-expand-1}~and~\eqref{eq:sir-expand-2} are satisfied. For $1\leq j\leq n$, and $1\leq i\leq j$, we construct $f(i,S(j))$ recursively, so that~\eqref{eq:sir-expand-1}~and~\eqref{eq:sir-expand-2} are satisfied. The base case follows from budget-balance, that is, $f(i,\{i\})=\alpha_{op}d(\{i\})$ for all $i\in\mathcal{N}$. Assume that for some $2\leq j\leq n$, we have defined $f(i,S(j-1))$ for all $1\leq i \leq j-1$. Then, we set
\begin{equation*}
\begin{split}
f(i,S(j)) &= f(i,S(j-1)) - \Big(\mathcal{IC}_i(S(j)) - \mathcal{IC}_i(S(j-1))\Big),\quad 1\leq i\leq j-1\\
f(j,S(j)) &= \mathcal{OC}(S(j)) - \sum_{i=1}^{j-1}f(i,S(j)).
\end{split}
\end{equation*}
By construction, it follows that~\eqref{eq:sir-expand-1} is satisfied, and $f$ is budget-balanced. It remains to be shown that~\eqref{eq:sir-expand-2} is also satisfied. By budget balance,
\begin{equation*}
\begin{split}
f(j,S(j)) &= \mathcal{OC}(S(j)) - \sum_{i=1}^{j-1}f(i,S(j))\\
&= \mathcal{OC}(S(j)) - \sum_{i=1}^{j-1}\Big(f(i,S(j-1)) - \Big(\mathcal{IC}_i(S(j)) - \mathcal{IC}_i(S(j-1))\Big)\Big)\\
&= \Big(\mathcal{OC}(S(j))-\mathcal{OC}(S(j-1))\Big) + \sum_{i=1}^{j-1}\Big(\mathcal{IC}_i(S(j)) - \mathcal{IC}_i(S(j-1))\Big)\\
&\leq \alpha_{op}d(\{j\}) - \mathcal{IC}_j(S(j)),
\end{split}
\end{equation*}
where, the last step follows from the assumption that~\eqref{eq:sir-feasible-constraints-alt} holds, and the previous step follows from the budget-balance property. This completes the proof.
\hfill\Halmos

\subsection{Proof of Theorem~\ref{thm:SIR-starvation-no-detour}}\label{appendix:proof-thm-SIR-starvation-no-detour}

First, we note that in the limit, when $\frac{\alpha_i}{\alpha_{op}}\rightarrow\infty$ for all $i\in\mathcal{N}$, the SIR-feasibility constraints~\eqref{eq:sir-feasible-constraints-single} reduce to
\begin{equation*}
S_{j-1}S_j + S_jD - S_{j-1}D \leq 0,\quad 2\leq j\leq n.
\end{equation*}
Since the points are from an underlying metric space, distances satisfy the triangle inequality, which means
\begin{equation*}
S_{j-1}S_j + S_jD - S_{j-1}D \geq 0,\quad 2\leq j\leq n.
\end{equation*}
Therefore, it must be that
\begin{equation*}
S_{j-1}S_j + S_jD - S_{j-1}D = 0,\quad 2\leq j\leq n.
\end{equation*}
By summing up the last $n-i$ equations, i.e., $i+1\leq j\leq n$, we get
\begin{equation*}
\sum_{j=i}^{n-1}S_jS_{j+1}+S_nD-S_iD = 0,
\end{equation*}
from which we obtain
\begin{equation*}
\gamma_r = \max_{i\in\mathcal{N}}\left(\frac{\sum_{j=i}^{n-1}S_jS_{j+1}+S_nD}{S_iD}\right) = 1.
\end{equation*}
This completes the proof.
\hfill\Halmos

\subsection{Proof of Theorem~\ref{thm:SIR-starvation-upper-1}}\label{appendix:proof-thm-SIR-starvation-upper-1}

First, we note that under the constraint $\frac{\alpha_i}{\alpha_{op}}\geq 1$ for all $i\in\mathcal{N}$, the SIR-feasibility constraints~\eqref{eq:sir-feasible-constraints-single} imply
\begin{equation}\label{eq:sir-feasible-constraints-single-eq}
S_{j-1}S_j + S_jD - S_{j-1}D \leq \frac{S_jD}{j},\quad 2\leq j\leq n.
\end{equation}
We begin by deriving an upper bound on the starvation factor of the $i$-th passenger, $1\leq i < n$, along any SIR-feasible route. (Note that, in any single dropoff instance, the starvation factor of the last passenger to be picked up is always $1$.) First, we sum up the last $n-i$ inequalities of~\eqref{eq:sir-feasible-constraints-single-eq}, i.e., $i+1\leq j\leq n$, to obtain
\begin{equation}\label{eq:partial-upper}
\sum_{j=i}^{n-1}S_jS_{j+1}+S_nD-S_iD \leq \sum_{j=i}^{n}\frac{S_jD}{j}.
\end{equation}
Next, we derive upper bounds for each $S_jD$, $i< j\leq n$, in terms of $S_iD$. The $j$-th SIR-feasibility constraint from~\eqref{eq:sir-feasible-constraints-single-eq} can be rewritten as
\begin{equation*}
S_jD - \frac{S_jD}{j} \leq S_{j-1}D - S_{j-1}S_j.
\end{equation*}
We know that $S_{j-1}S_j+S_{j-1}D\geq S_jD$, since all points are from an underlying metric space and therefore, distances are symmetric and satisfy the triangle inequality. Using this inequality above, we get
\begin{equation*}
\begin{split}
& S_jD - \frac{S_jD}{j} \leq S_{j-1}D - (S_jD - S_{j-1}D)\\
\Longrightarrow\;\; & (2j-1) S_jD \leq 2j S_{j-1}D\\
\Longrightarrow\;\; & S_jD \leq \frac{2j}{2j-1}S_{j-1}D.
\end{split}
\end{equation*}
Unraveling the recursion yields
\begin{equation*}
S_jD \leq \left(\prod_{k=i+1}^{j}\frac{2k}{2k-1}\right)S_iD = \frac{C_j}{C_i}S_iD,
\end{equation*}
where, for $m\geq 1$, $C_m = \prod_{k=1}^m\frac{2k}{2k-1}$. We can evaluate $C_j$ as follows:
\begin{equation*}
C_j = \prod_{k=1}^j\frac{2k}{2k-1} = \prod_{k=1}^j\frac{(2k)^2}{2k(2k-1)} = \frac{2^{2j}(j!)^2}{(2j)!} = \frac{2^{2j}}{\binom{2j}{j}}.
\end{equation*}
We then use a known lower bound for the central binomial coefficient, $\binom{2j}{j}\geq \frac{2^{2j-1}}{\sqrt{j}}$, to obtain $C_j\leq 2\sqrt{j}$. This yields $S_jD\leq \frac{2\sqrt{j}}{C_i}S_iD$. Substituting in~\eqref{eq:partial-upper}, we get
\begin{equation*}
\begin{split}
& \sum_{j=i}^{n-1}S_jS_{j+1}+S_nD-S_iD \leq \sum_{j=i}^{n}\frac{2}{C_i}\frac{\sqrt{j}}{j}S_iD = \frac{2}{C_i}\left(\sum_{j=i}^{n}\frac{1}{\sqrt{j}}\right)S_iD\\
\Longrightarrow\;\; & \sum_{j=i}^{n-1}S_jS_{j+1}+S_nD \leq \left(1+\frac{2}{C_i}\left(\sum_{j=i}^{n}\frac{1}{\sqrt{j}}\right)\right)S_iD.
\end{split}
\end{equation*}
This results in the desired upper bound for the starvation factor of the $i$-th passenger along any SIR-feasible route:
\begin{equation*}
\gamma_r(i) \leq 1+\frac{2}{C_i}\left(\sum_{j=i}^{n}\frac{1}{\sqrt{j}}\right).
\end{equation*}
The starvation factor of a route is the maximum starvation factor of all its passengers:
\begin{equation*}
\gamma_r = \max_{i\in\mathcal{N}}\gamma_r(i) \leq \max_{1\leq i < n} \left(1+\frac{2}{C_i}\left(\sum_{j=i}^{n}\frac{1}{\sqrt{j}}\right)\right) = 1+\frac{2}{C_1}\left(\sum_{j=1}^{n}\frac{1}{\sqrt{j}}\right) = 1+\sum_{j=1}^{n}\frac{1}{\sqrt{j}},
\end{equation*}
since $C_i$ is increasing in $i$ and $C_1=2$. The final step is to show that for all $n\geq 1$, $\sum_{j=1}^{n}\frac{1}{\sqrt{j}}\leq 2\sqrt{n}-1$. The proof is by induction. The base case (for $n=1$) is satisfied with equality. Assume that the statement is true for some $k\geq 1$. Then, for $k+1$, we have, $\sum_{j=1}^{k+1}\frac{1}{\sqrt{j}}\leq 2\sqrt{k}-1+\frac{1}{\sqrt{k+1}} = \frac{\sqrt{4k(k+1)}+1}{\sqrt{k+1}}-1 \leq \frac{\sqrt{4k(k+1)+1}+1}{\sqrt{k+1}}-1 = \frac{(2k+1)+1}{\sqrt{k+1}}-1 = 2\sqrt{k+1}-1$, which completes the inductive step. Using this bound, we get $\gamma_r \leq 2\sqrt{n}$, as desired. This completes the proof.
\hfill\Halmos

\subsection{Proof of Theorem~\ref{thm:SIR-starvation-upper-2}}\label{appendix:proof-thm-SIR-starvation-upper-2}

First, we note that in the limit, when $\frac{\alpha_i}{\alpha_{op}}\rightarrow 0$ for all $i\in\mathcal{N}$, the SIR-feasibility constraints~\eqref{eq:sir-feasible-constraints-single} reduce to
\begin{equation}\label{eq:sir-feasible-constraints-single-eq-2}
S_{j-1}S_j + S_jD - S_{j-1}D \leq S_jD,\quad 2\leq j\leq n.
\end{equation}
Our proof technique is exactly the same as that for Theorem~\ref{thm:SIR-starvation-upper-1}. We begin by deriving an upper bound on the starvation factor of the $i$-th passenger, $1\leq i < n$, along any SIR-feasible route, by summing up the last $n-i$ inequalities of~\eqref{eq:sir-feasible-constraints-single-eq-2} to obtain
\begin{equation}\label{eq:partial-upper-2}
\sum_{j=i}^{n-1}S_jS_{j+1}+S_nD-S_iD \leq \sum_{j=i}^{n}S_jD.
\end{equation}
Next, we derive upper bounds for each $S_jD$, $i< j\leq n$, in terms of $S_iD$. The $j$-th SIR-feasibility constraint from~\eqref{eq:sir-feasible-constraints-single-eq-2} can be rewritten as $S_{j-1}S_j \leq S_{j-1}D$. Using this in the triangle inequality $S_jD\leq S_{j-1}S_j+S_{j-1}D$, we get $S_jD \leq 2S_{j-1}D$. Unraveling this recursion then yields $S_jD \leq 2^{j-i}S_iD$. Substituting this in~\eqref{eq:partial-upper-2},
\begin{equation*}
\begin{split}
& \sum_{j=i}^{n-1}S_jS_{j+1}+S_nD-S_iD \leq \sum_{j=i}^{n}2^{n-i}S_iD = \sum_{j=0}^{n-i}2^jS_iD = \left(2^{n-i+1}-1\right)S_iD\\
\Longrightarrow\;\; & \sum_{j=i}^{n-1}S_jS_{j+1}+S_nD \leq 2^{n-i+1}S_iD.
\end{split}
\end{equation*}
Thus, the starvation factor of the $i$-th passenger along any SIR-feasible route is upper bounded as $\gamma_r(i) \leq 2^{n-i+1}$. Finally,
\begin{equation*}
\gamma_r = \max_{i\in\mathcal{N}}\gamma_r(i) \leq \max_{1\leq i < n} 2^{n-i+1} = 2^{n}.
\end{equation*}
This completes the proof.
\hfill\Halmos

\subsection{Proof of Theorem~\ref{thm:SIR-starvation-lower}}\label{appendix:proof-thm-SIR-starvation-lower}

To reduce notational clutter, we let $z_j=\left(1+\frac{1}{\alpha_{op}}\sum_{k=1}^{j-1}\alpha_k\right)^{-1}$, for $1\leq j\leq n$. We exhibit a single dropoff instance of size $n$ for which there is a unique SIR-feasible path whose starvation factor is exactly $\sum_{j=1}^{n}z_j$.

This instance is depicted in Fig.~\ref{fig:SIR-starvation-lower}. Here, $S_jD=\ell$ for $1\leq j\leq n$, and $S_{j-1}S_{k} > S_{j-1}S_{j}=z_j\ell$ for $2\leq j < k\leq n$. It is straightforward to see that the route $(S_1,S_2,\ldots,S_n,D)$ is SIR-feasible from~\eqref{eq:sir-feasible-constraints-single}, since for $2\leq j\leq n$, we have $S_{j-1}S_j + S_jD - S_{j-1}D = z_j\ell + \ell - \ell = z_j S_jD$, by construction. Thus, the starvation factor for this route is given by $\sum_{j=2}^{n}z_j+1 = \sum_{j=1}^{n}z_j$, as desired.

\begin{figure}[ht]
\centering
    \includegraphics[width=0.6\columnwidth]{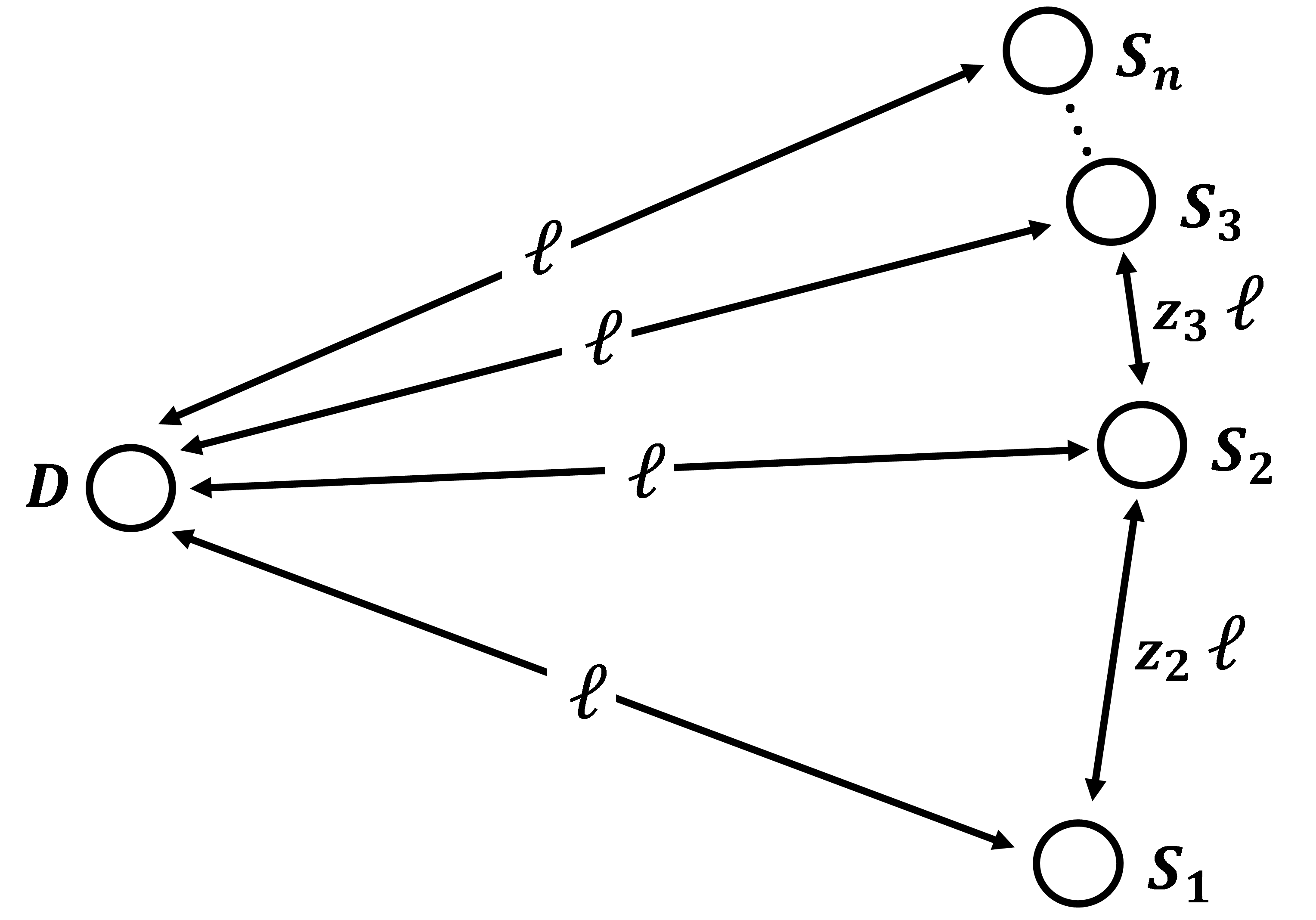}
        \caption{Single dropoff instance to establish lower bound on the SIR-starvation factor.\label{fig:SIR-starvation-lower}}
\end{figure}

It remains to be shown that no other route is SIR-feasible. First, we note that the SIR-feasibility constraints~\eqref{eq:sir-feasible-constraints-single} for this example simplify to
\begin{equation}\label{eq:sir-feasible-constraints-single-lower}
S_{j-1}S_j \leq z_j \ell,\quad 2\leq j\leq n,
\end{equation}
where $z_2>z_3>\ldots>z_n$, and $S_j$ refers to the $j$-th pickup point along the route. The proof is by induction. First, consider the pickup point $S_1$, whose distance from $S_2$ is $z_2\ell$, and from any other pickup point is strictly greater than $z_2\ell$, by construction. From~\eqref{eq:sir-feasible-constraints-single-lower}, it can be seen that no two pickup points that are more than $z_2\ell$ apart can be visited in succession, and that the only way to visit two pickup points that are exactly $z_2\ell$ apart is to visit them first and second. Thus, any SIR-feasible route must begin by visiting $S_1$ and $S_2$ first. This logic can be extended to build the unique SIR-feasible route that we analyzed above. This completes the proof.
\hfill\Halmos

\section{Proofs from Section~\ref{SEC:ALG-SIR}}\label{app:proofs-sec-6}

\subsection{Proof of Theorem~\ref{thm:sir-feasibility-hard}}\label{appendix:proof-thm-sir-feasibility-hard}

Given an instance of the Hamiltonian path problem in the form of a simple, undirected graph $G=(V,E)$, where $V=\{v_1,v_2,\ldots,v_n\}$, we construct an instance of \texttt{SIR-Feasibility} as follows. Let $P_j$ denote a pickup point corresponding to vertex $v_j\in V$. Let $\mathcal{N}=\{P_1,P_2,\ldots,P_n\}$ denote the set of pickup points, and $D$ denote the common dropoff point. Then, we set the pairwise distances to
\begin{equation*}
P_iP_j =
\begin{cases}
\frac{\ell}{n}, & (v_i,v_j)\in E\\
\ell, & otherwise,
\end{cases}
\end{equation*}
where $\ell>0$ is any constant. We also set $P_iD=\ell$ for all $i$, and $\alpha_{op}=\alpha_{1}=\alpha_{2}=\ldots=\alpha_{n}$, so that the SIR-feasibility constraints are given by~\eqref{eq:sir-feasible-constraints-single-eq}. Then, there is a one-to-one correspondence between the set of Hamiltonian paths in $G$ and the set of SIR-feasible routes in the corresponding instance of \texttt{SIR-Feasibility}, as follows:
\begin{enumerate}
\item Given a Hamiltonian path through a sequence of vertices $(u_1,u_2,\ldots,u_n)$ in $G$, let the corresponding sequence of pickup points be $(S_1,S_2,\ldots,S_n)$. Then, the route $(S_1,S_2,\ldots,S_n,D)$ is SIR-feasible, since the SIR-feasibility constraints~\eqref{eq:sir-feasible-constraints-single-eq} reduce to $S_{j-1}S_j\leq\frac{\ell}{j}$ for $2\leq j\leq n$, which are true, by construction.
\item Given an SIR-feasible route $(S_1,S_2,\ldots,S_n,D)$, let the corresponding sequence of vertices in $G$ be $(u_1,u_2,\ldots,u_n)$. Since the route is SIR-feasible, it must be that $S_{j-1}S_j\leq\frac{\ell}{j}$ for $2\leq j\leq n$. By construction, this means that $S_{j-1}S_j=\frac{\ell}{n}$, implying that $(u_{j-1},u_{j})\in E$ for $2\leq j\leq n$. Thus, the corresponding path is Hamiltonian.
\end{enumerate}
Hence, any algorithm for \texttt{SIR-Feasibility} can be used to solve the undirected Hamiltonian path problem with a polynomial overhead in running time. Since the latter is NP-hard, so is the former. This completes the proof.
\hfill\Halmos

\subsection{Proof of Theorem~\ref{thm:opt-sir-route-hard}}\label{appendix:proof-thm-opt-sir-route-hard}

Given an instance of \textit{Metric-Path-TSP} in the form of a complete undirected graph $G=(V,E)$ and distances $d(v_i,v_j)$ for each $v_i,v_j\in V$ from a metric space, we construct an instance of \texttt{Opt-SIR-Route} as follows. Let $P_j$ denote a pickup point corresponding to vertex $v_j\in V$. Let $\mathcal{N}=\{P_1,P_2,\ldots,P_n\}$ denote the set of pickup points, and $D$ denote the common dropoff point.

We set the pairwise distances $P_iP_j$ to be equal to $d(v_i,v_j)$ for all $v_i,v_j\in V$. We also set $P_iD=L$ for all $i$, where
\begin{equation*}
L > n\left(\max_{1\leq i<j\leq n}P_iP_j\right)
\end{equation*}
is any constant. We also set $\alpha_{op}=\alpha_{1}=\alpha_{2}=\ldots=\alpha_{n}$, so that the SIR-feasibility constraints are given by~\eqref{eq:sir-feasible-constraints-single-eq}. It is easy to see that for any route $(S_1,S_2,\ldots,S_n,D)$, these SIR-feasibility constraints reduce to $S_{j-1}S_j\leq\frac{\ell}{j}$ for $2\leq j\leq n$, which are true, by construction and our choice of $L$. Thus, all $n!$ routes in our constructed instance of \texttt{Opt-SIR-Route} are SIR-feasible. Moreover, by construction, the distance traveled along any route is exactly $L$ more than the weight of the path determined by the corresponding sequence of vertices in $G$. This implies that any optimal SIR-feasible route is given by a sequence of pickup points corresponding to an optimal Hamiltonian path in $G$, followed by a visit to $D$. Hence, any algorithm for \texttt{Opt-SIR-Route} can be used to solve \textit{Metric-Path-TSP} with a polynomial overhead in running time. Since the latter is NP-hard, so is the former. This completes the proof.
\hfill\Halmos

\subsection{Proof of Reduction from Section~\ref{ssec:opt-total-order-allocation}}\label{appendix:proof-reduction}

\begin{lemma}\label{lemma:0}
Any integral maximum flow from $s$ to $t$ must be comprised of $m'$ vertex-disjoint paths between the source vertex $s$ and the dropoff vertex $n+1$.
\end{lemma}

\proof{Proof.}

First, we observe that any integral \textit{feasible} flow from $s$ to $t$ in the network is comprised of vertex-disjoint paths between the source vertex $s$ and the dropoff vertex $n+1$, each carrying one unit of flow. This is because, every entry vertex $u_{in}$ has only one outgoing edge, namely, the one directed to its corresponding exit vertex $u_{out}$, which has unit capacity. (Similarly, every exit vertex only has one incoming edge, of unit capacity.) Thus, once a unit of flow is routed through $u_{in}$ and $u_{out}$ by some path, another path cannot route any additional flow through these vertices. Since the maximum flow on the network is $m'$ units, any integral feasible maximum flow would have to have $m'$ such vertex-disjoint paths between $s$ and $n+1$, each carrying one unit of flow. This completes the proof.
\hfill\Halmos

\begin{lemma}\label{lemma:1}
In any integral minimum cost flow, for every $u\in\mathcal{N}$, there is exactly one unit of flow entering $u_{in}$ and exactly one unit of flow leaving $u_{out}$.
\end{lemma}

\proof{Proof.}

From the proof of Lemma~\ref{lemma:0}, any integral feasible flow from $s$ to $t$ in the network is comprised of vertex-disjoint paths between the source vertex $s$ and the dropoff vertex $n+1$. Suppose by way of contradiction, an integral minimum cost flow does not route any flow through $v_{in}$ for some $v\in\mathcal{N}$. Let $\mathcal{G}_v$ denote the set of passengers $z\in\mathcal{N}$ such that $z<v$ and a unit of flow is routed via $\left(z_{in},z_{out}\right)$. Consider two cases:
\begin{enumerate}
\item \textbf{Case 1:} $\mathcal{G}_v\neq\emptyset$. Let $u=\max\mathcal{G}_v$, and let $\mathcal{P}_u$ be the path that carries a unit of flow from $s$ to $n+1$ through $u_{in}$ and $u_{out}$. The first vertex in $\mathcal{P}_u$ after $u_{out}$ is either an entry vertex $w_{in}$ for some $w\in\mathcal{N}$ (with $w>v$), or the dropoff vertex $n+1$. Then, we construct a new flow where $\mathcal{P}_u$ is modified to route its unit of flow from $u_{out}$ first to $v_{in}$ to $v_{out}$ and then to $w_{in}$ or $n+1$, as the case may be. (Note that this new flow is feasible, since $u<v<w<n+1$.) If $M$ and $M'$ denote the costs of the original flow and the new flow, then, we show that $M'<M$, contradicting the optimality of $M$:
    \begin{itemize}
    \item If the original flow took the route $u_{out}\rightarrow w_{in}$, and consequently, the new flow takes the route $u_{out}\rightarrow v_{in}\rightarrow v_{out}\rightarrow w_{in}$, then, $M' = M + S_{u}S_{v} - L + S_{v}S_{w}  - L - \left(S_{u}S_{w} - L\right) < M$ by our choice of $L$.
    \item If the original flow took the route $u_{out}\rightarrow n+1$, and consequently, the new flow takes the route $u_{out}\rightarrow v_{in}\rightarrow v_{out}\rightarrow n+1$, then, $M' = M + S_{u}S_{v} - L + S_{v}S_{n+1} - S_{u}S_{n+1} < M$ by our choice of $L$.
    \end{itemize}
\item \textbf{Case 2:} $\mathcal{G}_v=\emptyset$. Let $w\in\mathcal{N}$ be such that a unit of flow is routed from $s$ to $w_{in}$, $\mathcal{P}_w$ denoting the corresponding path. There may be more than one choice for $w_{in}$ as defined, but all of them satisfy $v<w$, since $\mathcal{G}_v=\emptyset$, so it does not matter which one is picked. As before, we construct a new flow where $\mathcal{P}_w$ is modified to route its unit of flow from $s$ first to $v_{in}$ to $v_{out}$ and then to $w_{in}$. (Note that this new flow is feasible, since $v<w$.) If $M$ and $M'$ denote the costs of the original flow and the new flow, $M' = M + S_{v}S_{w} - L < M$ by our choice of $L$, contradicting the optimality of $M$.
\end{enumerate}
This completes the proof.
\hfill\Halmos

\begin{lemma}\label{lemma:2}
The partition of $\mathcal{N}$ according to the $m'$ vertex-disjoint paths between $s$ and $n+1$ in an integral minimum cost maximum flow corresponds to the optimal allocation of the $n$ totally ordered passengers among $m'$ uncapacitated vehicles in the single dropoff scenario.
\end{lemma}

\proof{Proof.}
From Lemma~\ref{lemma:0} and Lemma~\ref{lemma:1}, we know that any integral minimum cost maximum flow $\mathcal{F}$ is comprised of $m'$ vertex-disjoint paths between $s$ and $n+1$ that cover all $n$ pickup points between them, by routing a unit of flow along $(u_{in},u_{out})$ for all $u\in\mathcal{N}$. We adopt a simplified representation of a path by removing the edges from the source vertex $s$, as well as the edges between $u_{in}$ and $u_{out}$, the entry and exit vertices corresponding to pickup points $u\in\mathcal{N}$. For example, a path $s\rightarrow u_{in}\rightarrow u_{out}\rightarrow v_{in}\rightarrow v_{out}\rightarrow n+1$ would be contracted to $u\rightarrow v\rightarrow n+1$. Note that this does not affect the cost computation, since only zero cost edges are removed. For any $u,v\in\mathcal{N}$, the cost of any edge $(u,v)$ in the new representation is simply the cost of the edge $(u_{out},v_{in})$ in the old representation. Similarly, for any $u\in\mathcal{N}$, the cost of any edge $(u,n+1)$ in the new representation is simply the cost of the edge $(u_{out},n+1)$ in the old representation. Let the set of these $m'$ paths be denoted as $\mathcal{P}_{\mathcal{F}}$. Thus, we have established a one-to-one correspondence between (a)~the set of all integral flows $\mathcal{F}$ comprised of $m'$ vertex-disjoint paths $\mathcal{P}_{\mathcal{F}}$ that collectively cover all $n$ pickup locations, and (b)~the set of all allocations of $n$ totally ordered passengers (traveling to a single dropoff location $n+1$) to $m'$ uncapacitated vehicles.

For any path $P\in\mathcal{P}_{\mathcal{F}}$, let $|P|$ denote the length of the path, that is, the number of edges in the path. The cost of path $P$ is then given by
\begin{equation*}
c(P) = \mathop{\sum\sum}_{\substack{1\leq u < v\leq n\\ (u,v)\in P}} (S_{u}S_{v}-L) + \sum_{\substack{1\leq u \leq n\\ (u,n+1)\in P}} S_{u}S_{n+1}.
\end{equation*}
Since all paths end with vertex $n+1$, there are $|P|-1$ terms in the first sum and $1$ term in the last sum. Thus, $c(P)$ can be equivalently written as
\begin{equation*}
c(P) = \mathop{\sum\sum}_{\substack{1\leq u < v\leq n+1\\ (u,v)\in P}} S_{u}S_{v} - (|P|-1)L.
\end{equation*}
The cost of flow $\mathcal{F}$ is simply the sum of the costs of the paths in $\mathcal{P}_{\mathcal{F}}$, given by
\begin{equation*}
c(\mathcal{F}) = \sum_{P\in\mathcal{P}_{\mathcal{F}}} c(P) = \mathop{\sum\sum}_{\substack{1\leq u < v\leq n+1\\ (u,v)\in\ \bigcup\mathcal{P}_{\mathcal{F}}}} S_{u}S_{v} - \sum_{P\in\mathcal{P}_{\mathcal{F}}}(|P|-1)L.
\end{equation*}
Since $|P|$, the length of path $P$, also denotes the number of pickup points covered by $P$, and all the $m'$ paths are vertex-disjoint (except for $n+1$), the summation in the second term is simply $n-m'$, independent of the flow $\mathcal{F}$. Thus,
\begin{equation}\label{eq:cost-of-flow}
c(\mathcal{F}) = \mathop{\sum\sum}_{\substack{1\leq u < v\leq n+1\\ (u,v)\in\ \bigcup\mathcal{P}_{\mathcal{F}}}} S_{u}S_{v} - (n-m')L = c(\mathcal{A}_{\mathcal{F}}) - (n-m')L,
\end{equation}
where $c(\mathcal{A}_{\mathcal{F}})$ denotes the cost (total vehicle-miles traveled) of the corresponding allocation of $n$ totally ordered passengers (traveling to a single dropoff location $n+1$) to $m'$ uncapacitated vehicles. From~\eqref{eq:cost-of-flow}, it is clear that the set of integral minimum cost maximum flows $\mathop{\arg\min}_{\mathcal{F}}c(\mathcal{F})$ also corresponds to the set of optimal allocations of $n$ totally ordered passengers among $m'$ uncapacitated vehicles in the single dropoff scenario. This completes the proof.
\hfill\Halmos

\begin{theorem}\label{thm:polytime-vrp}
There exists a poly-time algorithm to find an optimal allocation of totally ordered passengers to uncapacitated vehicles in the single dropoff scenario.
\end{theorem}

\proof{Proof.}
Using the one-to-one correspondence established in Lemma~\ref{lemma:2}, for each ``guess'' $1 \leq m'\leq n$, we find the corresponding optimal allocation by solving a minimum cost maximum flow problem in poly-time, finally choosing a guess with the overall least cost allocation.
\hfill\Halmos

\end{APPENDICES}  


\bibliographystyle{informs2014}
\bibliography{xharecost-msom}
\end{document}